\documentclass[pra,aps,footinbib,twocolumn,superscriptaddress,citeautoscript]{revtex4-1}

\usepackage[utf8x]{inputenc}
\usepackage{array}
\usepackage{amssymb}
\usepackage{amsmath}
\usepackage{amsfonts}
\usepackage{color}
\usepackage{graphicx}
\usepackage{bm}
\usepackage{dsfont}
\usepackage{url}
\usepackage{braket}
\usepackage{tikz}
\usepackage{refcount}
\usepackage[unicode]{hyperref}
\hypersetup{
   unicode=true,          
   plainpages=false,
   colorlinks=true,       
   citecolor=blue,        
}

\graphicspath{{.}{./EPS/}}

\newcommand* {\vek}[1]{{\ensuremath{\bm{\mathrm{#1}}}}}

\newcommand* {\kk}{\vek{k}}

\newcommand{\rme}{\mathrm{e}}
\newcommand{\rmi}{\mathrm{i}}
\newcommand{\x}{{\mathrm{x}}}
\newcommand{\y}{{\mathrm{y}}}
\newcommand{\z}{{\mathrm{z}}}
\newcommand{\absD}{{\left| \Delta \right|}}
\newcommand{\absB}{{\left| \B^\mathrm{H} \right|}}
\newcommand{\absE}{{\left|E_{\k,\pm}\right|}}
\newcommand{\absEa}{{\left|E_{\k,\alpha}\right|}}

\renewcommand{\k}{{\mathbf{k}}}
\newcommand{\B}{{\mathbf{B}}}
\newcommand{\kp}{{\mathbf{k}^\prime}}
\renewcommand{\r}{\mathbf{r}}
\newcommand{\ua}{\uparrow}
\newcommand{\da}{\downarrow}
\renewcommand{\d}{\ensuremath{\mathrm{d}}}

\makeatletter
\newcommand\footnoteref[1]{\protected@xdef\@thefnmark{\ref{#1}}\@footnotemark}
\makeatother

\begin{document}
\title{Vortex dipole precession in a Fermi superfluid with Rashba spin-orbit coupling}
\author{L. A. Toikka}
\affiliation{Center for Theoretical Physics of Complex Systems, Institute for Basic Science (IBS), Daejeon 34051, Republic of Korea}
\date{\today}
\begin{abstract}
We present a closed-form expression for the time dynamics of a balanced Fermi superfluid with $s$-wave pairing and Rashba spin-orbit (SO) coupling. Solving the associated self-consistency conditions for an initial number density given by a Gaussian and an initial gap containing a vortex dipole, we show that Rashba SO coupling results in precession of the vortex dipole.  The integrated self-consistent gap decays as a function of time, characterising a `transient Fermi superfluid'. Our analytic solution forms a starting point for studies of vortex dynamics -- in particular braiding of Majorana fermions -- in Fermi superfluids in the presence of spin population imbalance and two-dimensional SO coupling, which are both required to realize a topological atomic Fermi gas in experiments.\looseness=-1
\end{abstract}
\maketitle

\textit{Introduction}
Ranging from high-temperature superconductors to neutron stars, attractively interacting Fermi systems are physically ubiquitous~\cite{leggett2006quantum}. In experiments~\cite{PhysRevLett.118.123401,PhysRevLett.92.150402,Yefsah2013}, where the densities and temperatures are low, the scattering between the fermions takes place typically in the $s$-wave channel
~\cite{PhysRevLett.93.090404}. However, the Pauli exclusion principle prevents scattering between identical fermions, which means that we require two or more different species of fermions to form the pairs. We call the additional quantum number `spin', and take it to be $1/2$.\looseness=-1

A natural choice for the spin quantization axis is given by the momentum of the fermions, a so-called local spin-frame. While the underlying system interacts through $s$-wave scattering, this results in a spin-mixing representation where the Fermi superfluid has an effective $p$-wave symmetry that is scaled by the inverse magnitude of the momentum~\cite{PhysRevLett.107.195305}. This is similar to the effective $p$-wave pairing that emerges due to spin-orbit (SO) coupling~\cite{Zhang2008}. However, the treatment of SO coupling becomes easier in the local spin-frame, and we show how the mean-field time dynamics can be thus solved analytically for a SO coupled but spin-balanced Fermi superfluid.

Experimentally, the atomic Fermi gas is a highly flexible quantum many-body system whose interactions, spin balance, and trapping geometries can be tuned nearly arbitrarily~\cite{Zwierlein05a,PhysRevLett.116.045304}. In the mean-field picture, it is possible to achieve band inversion leading to a topological phase~\cite{1367-2630-13-6-065004} by combining the effects of two-dimensional SO coupling~\cite{PhysRevLett.109.095302,PhysRevLett.109.095301}, and spin imbalance~\cite{Zwierlein27012006,Partridge27012006}. Recently, highly tunable 2D SO coupling was demonstrated experimentally~\cite{Huang16}. These advances open up a promising perspective for creating an atomic topological superfluid~\cite{PhysRevA.85.033622,PhysRevA.86.063604} in the laboratory in the near future.

When the Fermi gas is in the topologically non-trivial phase, it is well-known that vortex cores house Majorana fermions~\cite{Wilczek09,PhysRevLett.100.027001,silaev14}. The Majorana vortices obey non-Abelian exchange statistics~\cite{PhysRevLett.86.268}, which means that they can be used to realize braiding operations. In the field of topological quantum computation~\cite{RevModPhys.80.1083}, quantum information is encoded in a non-local way using the degeneracy of the zero-energy Majorana fermions. One ingredient needed for performing the quantum logic operations is non-commutative braiding, which amounts to exchanging two Majorana vortices adiabatically~\cite{0268-1242-27-12-124003}. Despite the theoretically promising topological protection and non-local spreading of the quantum information to reduce decoherence effects, it is less well understood how to achieve braiding of Majorana fermions in such a topological system experimentally~\cite{aliceanp2011}. While considering a non-topological atomic Fermi gas, we demonstrate that the exchange of vortices can take place as a result of Rashba SO coupling. This result is a promising first step towards possible new ways to achieve Majorana braiding operations by using the experimental flexibility and advantages of ultra-cold atoms. 

Quantised vortices are hallmarks of superfluidity~\cite{Zwierlein2005}, and understanding their properties is an important general probe for the underlying physics of the system. Focussing on a vortex-dipole initial state in the superfluid gap and an initial number density given by a Gaussian, we show that Rashba SO coupling results in precession of the vortex dipole. The rotation and frequency of the precession is determined by the sign and magnitude of the 2D SO coupling, respectively. Significantly, precession of a vortex dipole in a topological Fermi superfluid amounts in principle to braiding operations performed on the Majorana fermions, which are localised at the vortex cores. Therefore, understanding the physics of vortex dipole precession in such exotic superfluids is important, for example, for constructing systems where the braiding can be precisely controlled by adjusting the Rashba SO coupling.

\textit{Mean-field description}
 The particular physical system we consider here is a two-dimensional Fermi gas with a spin-singlet $s$-wave pairing gap, represented by the complex number $\Delta(\k,t)$, under Rasbha SO coupling $\lambda_\kk = \lambda(k_\x + \rmi k_\y)$~\cite{Manchon15} and Zeeman-induced spin-population imbalance $h$. We do not consider spin-triplet pairing. To model our system, we start with the mean-field Bogoliubov-de Gennes (BdG) Hamiltonian~\cite{Jiang2011,Zhou2011,Liu2012b}
\begin{equation}
\begin{split}
\mathcal{H}(t) &= \int \d^2 \textbf{r}\, \left[ \sum_{\sigma, \sigma^\prime} \hat{\psi}_\sigma^\dagger (\textbf{r},t) \hat{K}_{\sigma \sigma^\prime}(\textbf{r},t) \hat{\psi}_{\sigma^\prime} (\textbf{r},t) \right. \\
&\left. + \Delta(\r,t)\hat{\psi}_\uparrow^\dagger (\textbf{r},t)\hat{\psi}_\downarrow^\dagger (\textbf{r},t) + \Delta^*(\r,t)\hat{\psi}_\downarrow (\textbf{r},t)\hat{\psi}_\uparrow (\textbf{r},t) + \text{h.c.}\right]\\
&= \frac{1}{\Omega}\sum_\kk \, \begin{pmatrix} \textbf{c}_\kk^\dagger & \textbf{c}_{-\kk} \end{pmatrix} H(\kk,t) \begin{pmatrix} \textbf{c}_\kk & \textbf{c}_{-\kk}^\dagger \end{pmatrix}^\mathrm{T}
\end{split}
\end{equation}  
with $ \textbf{c}_\kk^\dagger = \begin{pmatrix}
c_{\k,\ua}^\dagger & c_{\k,\da}^\dagger
\end{pmatrix}$ where $c_{\k,\sigma}^\dagger = \frac{1}{\sqrt{\Omega}} \int \d^2 \textbf{r}\, \rme^{\rmi \kk \cdot \textbf{r}}\,  \hat{\psi}_{\sigma}^\dagger (\textbf{r},t)$, where $\Omega$ is the volume occupied by the system, $\hat{K}_{\sigma \sigma^\prime}$ contains the SO coupling and free dispersion, and the operator-valued field $\hat{\psi}_{\sigma}^\dagger (\textbf{r},t)$ creates a spin-$\sigma$ ($\sigma = \uparrow, \downarrow$) fermion at location $\textbf{r}$ at time $t$. The mean-field Hamiltonian matrix is given by
\begin{equation}
\label{eqn:HswaveFermiSF-p}
\begin{split}
H(\kk) &= \begin{pmatrix}
H_0 - h \sigma_\mathrm{z} + \lambda \textbf{g}_\kk \cdot \boldsymbol{\sigma} & -\rmi \, \Delta(\k,t) \sigma_\mathrm{y} \\
\rmi \, \Delta(\k,t)^* \sigma_\mathrm{y} & -H_0 + h \sigma_\mathrm{z} + \lambda \textbf{g}_\kk \cdot \boldsymbol{\sigma}^*  \\
\end{pmatrix},
\end{split}
\end{equation}
where $\kk = \begin{pmatrix} k_\x &  k_\y \end{pmatrix}^\mathrm{T}$ is the momentum, 
$H_0 = \left(\epsilon_\kk - \bar{\mu}\right) \sigma_0$ with the single-particle dispersion $\epsilon_\kk = \hbar^2\kk^2/(2m)$, average chemical potential $\bar{\mu} = \left( \mu_\uparrow + \mu_\downarrow \right) / 2$ and Zeeman field $h = \left( \mu_\uparrow - \mu_\downarrow \right) / 2$, 
$\sigma_0$ is the identity matrix, 
$\boldsymbol{\sigma} =  \begin{pmatrix} \sigma_\x &  \sigma_\y \end{pmatrix}^\mathrm{T}$ together with $\sigma_\z$ are the Pauli matrices, 
$\textbf{g}_\kk = \sigma_\z \kk  =  \begin{pmatrix} k_\x &  -k_\y \end{pmatrix}^\mathrm{T}$, 
and $\lambda$ represents the strength of the Rashba SO coupling. The Nambu spinor basis for the Hamiltonian~\eqref{eqn:HswaveFermiSF-p} is given by $\begin{pmatrix}
u_{\k\eta,\ua}(t) &
u_{\k\eta,\da}(t)& v_{-\k\eta,\ua}(t)& v_{-\k\eta,\da}(t)\end{pmatrix}^\mathrm{T}$ where $\eta$ enumerates the bands.

At zero temperature the self-consistency conditions for the \textit{s}-wave pair potential and
particle-number density read~\cite{deGennes1989,ketterson1999superconductivity} (Appendix~\ref{sec:appselfc})
$\Delta(\k,t) = \frac{g}{\Omega}\left( \sum_{E_\eta<0}  u_{\k\eta,\ua} v_{-\k\eta,\da}^* + \sum_{\k,E_\eta>0}  u_{-\k\eta,\da} v_{\k\eta,\ua}^* \right)$ and $n_\sigma(\k,t) = \frac{1}{\Omega}\left[\sum_{E_{\eta}<0} \left|
u_{\k\eta,\sigma} \right|^2 + \sum_{E_{\eta}>0} \left|
v_{\k\eta,\sigma} \right|^2\right]$ respectively, where $g < 0$ is the strength of the short-range contact potential between the fermions. Here  $-E_2 = E_3 = \left|E_{\k,-}\right|$ and $-E_1 = E_4 = \left|E_{\k,+}\right|$, where $E_{\k,\pm}$ are the eigenvalues of $H(\k)$, which follows from the particle-hole symmetry $u_{\k\eta_1,\sigma} \stackrel{E_{\eta_1}=-E_{\eta_2}}{=} v_{\k\eta_2,\sigma}^*$~\footnote{In our case, the particle-hole symmetry is defined by $\{ \Xi_\mathrm{s-wave}, H(\k) \} = 0$, where $\Xi_\mathrm{s-wave} = \rme^{\rmi \theta} \tau_\x \otimes \sigma_0 K$, where $K$ is the complex conjugation operator that also flips the sign of the momentum in momentum space, and $\theta$ is arbitrary. A direct calculation verifies this: $\Xi_\mathrm{s-wave} H(\k)\Xi_\mathrm{s-wave}^{-1} = \rme^{\rmi \theta}\tau_\x \otimes \sigma_0 K \left[H(\k) \left(\rme^{\rmi \theta}\tau_\x \otimes \sigma_0 K\right)^{-1} \right]  = \rme^{\rmi \theta}\tau_\x \otimes \sigma_0 H^*(-\k) KK^{-1} \rme^{-\rmi \theta}\left(\tau_\x \otimes \sigma_0 \right)^{-1}  = \tau_\x \otimes \sigma_0 H^*(-\k) \left(\tau_\x \otimes \sigma_0 \right)^{-1} = -H(\k)$. Here we have for clarity defined $\tau_\x$ to be the Pauli matrix in the particle-hole space, while $\sigma_0$ is the identity matrix in spin space.}. The Hamiltonian $H(\k)$ has time-reversal symmetry~\footnote{In our case, the time-reversal symmetry is defined by $\left[ T, H(\k) \right] = 0$, where $T = \rme^{\rmi \theta} \tau_0 \otimes \sigma_\y K$ ($T^2 = -1$, for half-integer spin), where $K$ is the complex conjugation operator that also flips the sign of the momentum in momentum space, and $\theta$ is arbitrary. Here we have for clarity defined $\tau_0$ to be the identity matrix in the particle-hole space, while $\sigma_\y$ is the Pauli matrix in spin space.} if $\Delta(\k,t) \in \mathbb{R}$ and $h = 0$.

\textit{Helicity basis (local spin-frame)}
We will now show how to reduce in an exact fashion the treatment of Rashba SO coupling into an easier problem when $h = 0$. For convenience let us define an angle $\varphi_\k = \mathrm{arg}\left(k_\x + \rmi k_\y \right)$ such that $k \rme^{\pm \rmi \varphi_\k} = k_\x \pm \rmi k_\y$. We define the `helicity basis' by
\begin{equation}
\label{eqn:HelBasisTF}
\begin{pmatrix}
h_{\k,+}\\
h_{\k,-}
\end{pmatrix}
= \frac{1}{\sqrt{2}}
\begin{pmatrix}
1 & \rme^{\rmi \varphi_\k}\\
\rme^{-\rmi \varphi_\k} & -1
\end{pmatrix}
\begin{pmatrix}
c_{\k,\ua}\\
c_{\k,\da}
\end{pmatrix}.
\end{equation}
The `helicity' label $\pm$ means that the spin direction is either parallel ($+$) or anti-parallel ($-$) to the in-plane momentum $\k$. In the helicity basis we write $\mathcal{H} = \sum_\kk \, \begin{pmatrix} \textbf{h}_\kk^\dagger & \textbf{h}_{-\kk} \end{pmatrix} H^\mathrm{H}(\kk) \begin{pmatrix} \textbf{h}_\kk & \textbf{h}_{-\kk}^\dagger \end{pmatrix}^\mathrm{T}$ with $\textbf{h}_\k = \begin{pmatrix}
h_{\k,+} & h_{\k,-}
\end{pmatrix}$, and the Hamiltonian matrix reads (Appendix~\ref{sec:app_HelTF}) 
\begin{equation}
\label{eqn:HswaveFermiSF-p-HelTF}
H^\mathrm{H}(\k) = \begin{pmatrix}
 H_0 + \lambda k \sigma_\mathrm{z} - \frac{h}{k} \textbf{g}_\k \cdot \boldsymbol{\sigma } & \frac{\Delta}{k}\left(k_\x \sigma_0 + \rmi k_\y \sigma_\z \right) \\
 \frac{\Delta^*}{k}\left(k_\x \sigma_0 - \rmi k_\y \sigma_\z \right)  & -H_0 - \lambda k \sigma_\mathrm{z} - \frac{h}{k} \textbf{g}_\k \cdot \boldsymbol{\sigma}^*
\end{pmatrix} .
\end{equation}
We denote the Nambu spinor basis for the Hamiltonian matrix~\eqref{eqn:HswaveFermiSF-p-HelTF} by 
$\begin{pmatrix}
u_{\k,\eta}^+(t) &
u_{ \k,\eta}^-(t)& v_{\k,\eta}^+(t)& v_{\k,\eta}^-(t)\end{pmatrix}^\mathrm{T}$.

Comparing to the representation $H(\k)$ of $\mathcal{H}$ in the original basis given in Eq.~\eqref{eqn:HswaveFermiSF-p}, in addition to the order parameter now having an effective $p$-wave symmetry in the helicity basis representation, the Zeeman field $h$ has been interchanged with the spin-orbit coupling $\lambda$. While pairing in the underlying physical system remains strictly $s$-wave because $\Delta(\k,t)$ is a complex number, the form of the Hamiltonian matrix in the helicity basis indicates that effective pairing exists only inside a given helicity sector, and that the pairing of helicity $\pm$ atoms has $\left( k_\x \pm \rmi k_\y \right)/k$ symmetry. This result allows us to decouple the $4\times 4$ Hamiltonian $H^\mathrm{H}(\k)$ into two $2\times 2$ Hamiltonians, one for each helicity subspace. An exact decoupling into helicity subspaces is possible only if $h = 0$, and the effect of a small $h$ could be considered perturbatively. Importantly, SO coupling can be retained in the helicity basis. 

We now focus on the spin-balanced Fermi superfluid. The two subspace Hamiltonians then read
\begin{equation}
\label{eqn:matH2x2pm}
H_\pm^\mathrm{H} = \begin{pmatrix}
\epsilon_\kk - \bar{\mu} \pm \lambda k & \frac{\Delta}{k} \left(k_\x \pm \rmi k_\y \right) \\
\frac{\Delta^*}{k} \left(k_\x \mp \rmi k_\y \right)& -\left(\epsilon_\kk - \bar{\mu} \pm \lambda k\right)
\end{pmatrix},
\end{equation}
and the corresponding Bogoliubov-de Gennes (BdG) equations are
\begin{subequations}
\label{eqn:BdGeq}
\begin{align}
\left(\epsilon_\kk - \bar{\mu} \pm \lambda k \right)u_{\k,\eta}^\pm + \frac{\Delta}{k} \left(k_\x \pm \rmi k_\y \right)v_{\k,\eta}^\pm &= \rmi \hbar \frac{\partial u_{\k,\eta}^\pm}{\partial t},\\ 
\label{eqn:BdGeq-v}
-\left(\epsilon_\kk - \bar{\mu} \pm \lambda k \right)v_{\k,\eta}^\pm + \frac{\Delta^*}{k} \left(k_\x \mp \rmi k_\y \right)u_{\k,\eta}^\pm &= \rmi \hbar \frac{\partial v_{\k,\eta}^\pm}{\partial t}.
\end{align}
\end{subequations}
In the helicity basis, at $T = 0$, the self-consistency equations are a sum of four terms (Appendix~\ref{sec:appselfc}),
\begin{subequations}
\label{eqn:SCHelicityBasis}
\begin{align}
\label{eqn:GapEqHelicityBasis}
\Delta(\k,t) &= 
-\frac{g}{\Omega} \sum_{\stackrel{E_\eta<0,}{\alpha=\pm}}\left( 
\rme^{-\alpha\rmi \varphi_\k} u_{\k,\eta}^\alpha \bar{v}_{-\k,\eta}^\alpha
+
u_{\k,\eta}^{-\alpha} \bar{v}_{-\k,\eta}^\alpha
\right),\\
\label{eqn:NumEqHelicityBasis}
n(\k,t) &= \frac{1}{2\Omega} \sum_{\alpha=\pm}\left(   \sum_{E_\eta>0} |v_{\k\eta}^\alpha |^2  +\sum_{E_\eta<0}  |u_{\k\eta}^\alpha |^2\right),
\end{align}
\end{subequations}
where the bar denotes complex conjugation. Focussing on the low-energy spectrum $\epsilon_\k \approx 0$ and considering $\Delta(\k,t) \in \mathbb{R}$, the subspace Hamiltonians~\eqref{eqn:matH2x2pm} are algebraically equivalent to the celebrated Dirac Hamiltonian $H_\mathrm{D} = m^*c^2 \sigma_\z + \hbar c \boldsymbol{\sigma} \cdot \k$ in two dimensions. In this case, the SO coupling gives rise to a linear-in-$k$ effective mass $m^*  =\left(\pm \lambda k - \bar{\mu}\right)/c^2$, an effective speed of light $ \hbar c = \Delta(\k,t)/k$, and the Pauli matrices $\boldsymbol{\sigma}$ represent real spin. The case of graphene~\cite{PhysRevLett.53.2449,PhysRevB.29.1685} can be obtained for vanishing effective mass $m^* = 0$, and the effective speed of light is given by the Fermi velocity: $\Delta(\k,t)/k = \hbar v_\mathrm{F}$.

\textit{Solution of the BdG equations}
Denoting $\phi^\pm_\k(t) = \begin{pmatrix}
u_{\k,\eta}^\pm(t) & v_{\k,\eta}^\pm(t)
\end{pmatrix}^\mathrm{T} 
$, we obtain from the system~\eqref{eqn:BdGeq} the time-evolution problem
\begin{equation}
\label{eqn:time-evprob}
\rmi \hbar \frac{\partial \phi^\pm_\k(t)}{\partial t}  = H^\mathrm{H}_\pm \phi^\pm_\k(t),
\end{equation}
which can be solved formally by writing $\phi^\pm_\k(t) = \rme^{-\frac{\rmi}{\hbar} \int^t  H^\mathrm{H}_\pm \,\d t } \phi^\pm_\k(0)$. We now provide an exact analytical solution for $\phi^\pm_\k(t)$ when $\{ \lambda, \Delta(\k), \bar{\mu}(\k)\}$ do not depend on time. We assume for small $\lambda \neq 0$ that our solution is still accurate enough to capture the dipole precession at least qualitatively.

Let us consider a single band $\eta_0$. In momentum space, we can evaluate the time dynamics for the Bogoliubov amplitudes exactly (Appendix~\ref{sec:app_BdGeval}),
\begin{subequations}
\label{eqn:BdGSolnuv-momspace}
\begin{align}
\label{eqn:BdGSolnuv-u-momspace}
&u^\pm_{\k,\eta_0}(t) = - \rmi  \sin{\left(\frac{\absE t}{\hbar}  \right)} \frac{\Delta \rme^{\pm\rmi \varphi_\k }}{\absE} \chi^\pm_0(\k)+\\
&\notag \left[\cos{\left(\frac{\absE t}{\hbar}  \right)} - \rmi  \sin{\left(\frac{\absE t}{\hbar}  \right)} \frac{\left(\epsilon_\k - \bar{\mu} \pm \lambda k \right) }{\absE}\right]\phi_0^\pm(\k),\\
\label{eqn:BdGSolnuv-v-momspace}
&v^\pm_{\k,\eta_0}(t) = \left[ - \rmi  \sin{\left(\frac{\absE t}{\hbar}  \right)} \frac{\Delta^* \rme^{\mp \rmi \varphi_\k} }{\absE}\right]\phi_0^\pm(\k)+\\
&\notag \left[\cos{\left(\frac{\absE t}{\hbar}  \right)}   - \rmi  \sin{\left(\frac{\absE t}{\hbar}  \right)} \frac{-\left(\epsilon_\k - \bar{\mu} \pm \lambda k \right) }{\absE}\right] \chi_0^\pm(\k).
\end{align}
\end{subequations}
Here $E_{\k,\pm} = \sqrt{\left(\epsilon_\k - \bar{\mu} \pm \lambda k \right)^2  + \absD^2 }$. Equation~\eqref{eqn:BdGSolnuv-momspace} is a closed-form exact solution for $\phi^\pm_\k(t)$ with the (arbitrary) initial condition $\phi^\pm_\k(0) = \begin{pmatrix}
\phi^\pm_0(\k) &  \chi^\pm_0(\k) \end{pmatrix}^\mathrm{T}$, and directly satisfies the Bogoliubov-de Gennes equations~\eqref{eqn:BdGeq}. We ignore the complication arising in taking the time derivatives in Eq.~\eqref{eqn:BdGeq} when $\{ \lambda, \Delta, \bar{\mu}\}$ depend on time. Using the inverse helicity transformation [Eqs.~\eqref{eqn:HelBasisTF1},~\eqref{eqn:heltfdefd} and~\eqref{eqn:heltfdefe}], Eq.~\eqref{eqn:BdGSolnuv-momspace} reads $\begin{pmatrix}
u_{\k, \ua} \\
u_{\k, \da} \\
\end{pmatrix} = \mathcal{T}_\k \begin{pmatrix}
u_{\k,\eta_0}^+ \\
u_{\k,\eta_0}^- \\
\end{pmatrix}$ and $\begin{pmatrix}
v_{\k, \ua} \\
v_{\k, \da} \\
\end{pmatrix} = \mathcal{T}^\mathrm{T}_{\k} \begin{pmatrix}
v_{\k,\eta_0}^+ \\
v_{\k,\eta_0}^- \\
\end{pmatrix}$ in the original representation.

\begin{figure}[t]
\begin{tikzpicture}
\pgfmathsetmacro{\rowad}{0.0}
\pgfmathsetmacro{\rowtwo}{-6.4}

 \node[] at (0-1.5, 0) { \includegraphics[width=0.3\textwidth,angle=0]{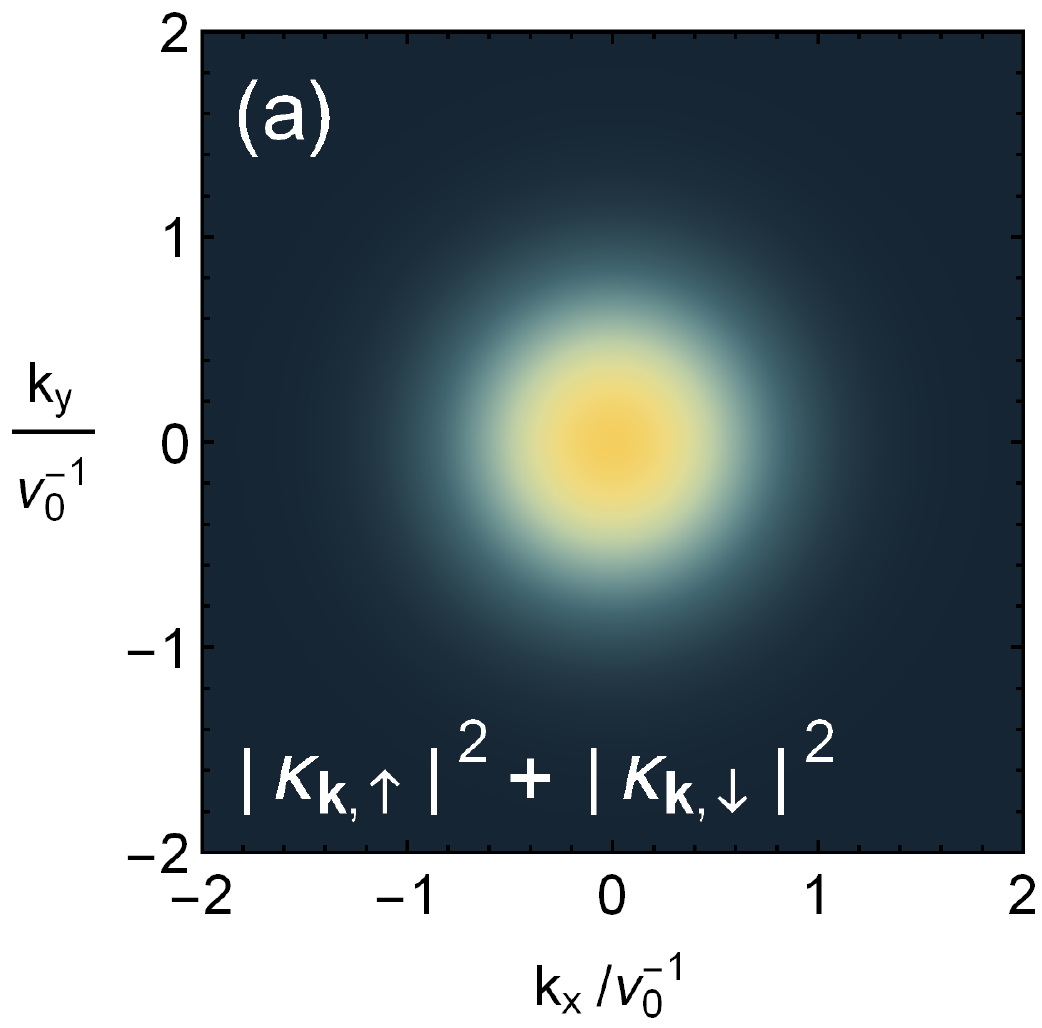}};
  \node[] at (0.7, 0.4) { \includegraphics[width=0.04\textwidth,angle=0]{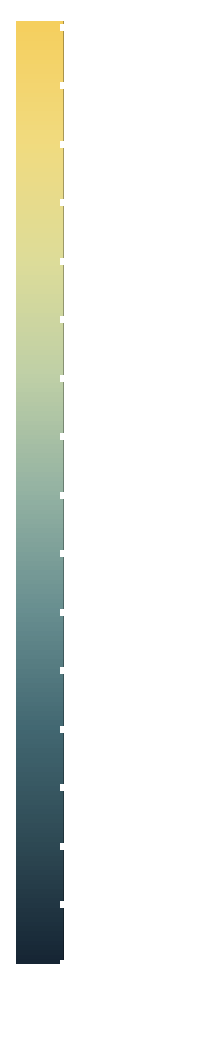}};
 \node[] at (4-1.5, 1.5) { \includegraphics[width=0.12\textwidth,angle=0]{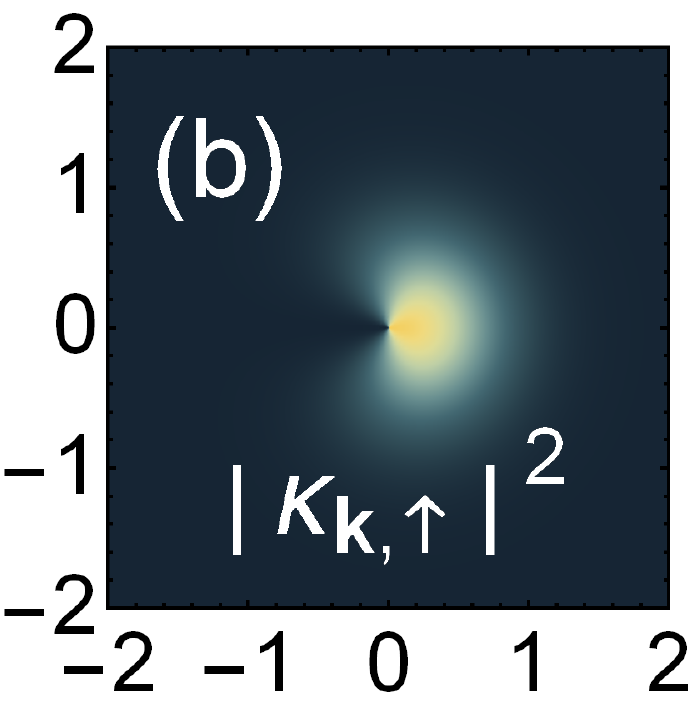}};   
 \node[] at (4-1.5, -1.0) { \includegraphics[width=0.12\textwidth,angle=0]{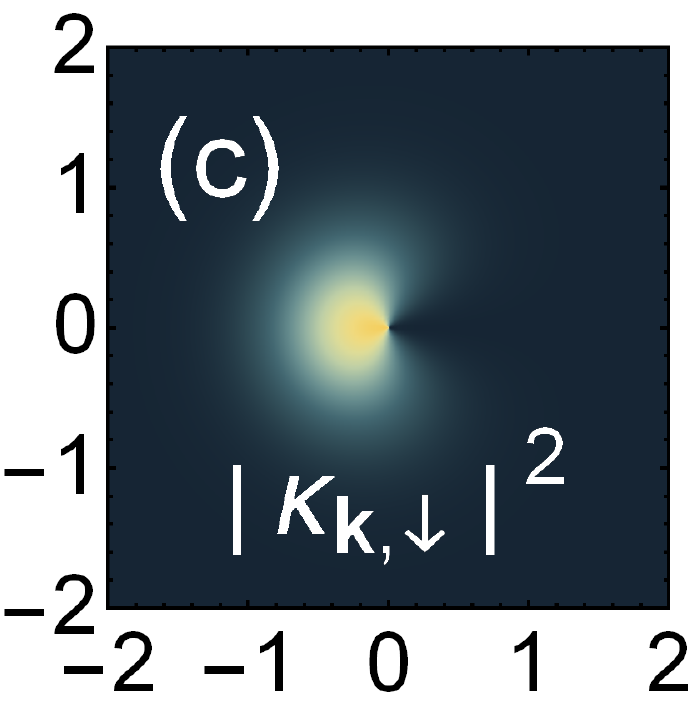}};
    
 \node[] at (0-2.8, -4.0+\rowad) { \includegraphics[width=0.13\textwidth,angle=0]{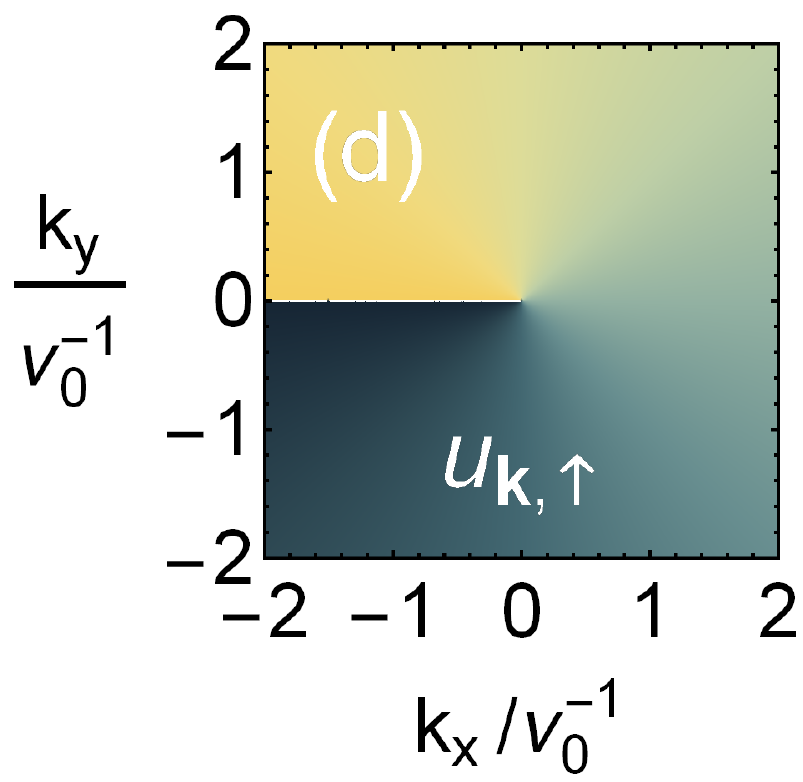}};   
  \node[] at (1.9-2.5, -4.0+\rowad) { \includegraphics[width=0.106\textwidth,angle=0]{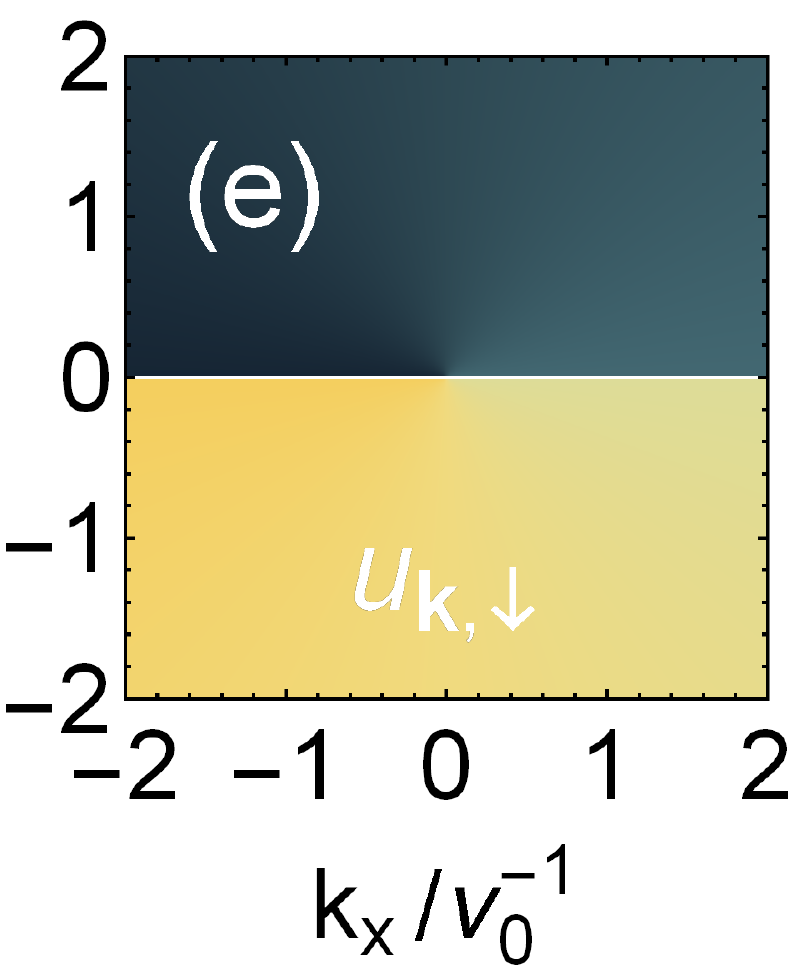}};   
   \node[] at (3.8-2.5, -4.0+\rowad) { \includegraphics[width=0.106\textwidth,angle=0]{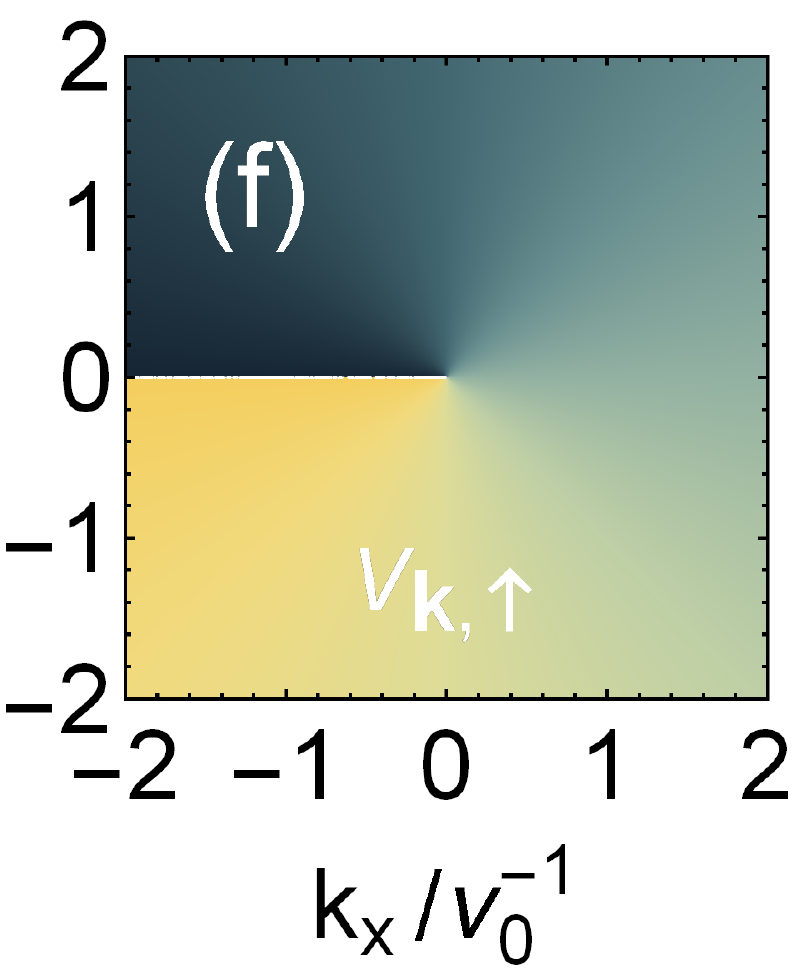}};   
   \node[] at (5.8-2.5, -2.7+\rowad) { \includegraphics[width=0.085\textwidth,angle=0]{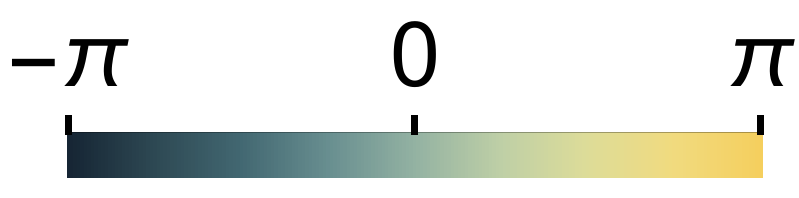}};
    \node[] at (5.7-2.5, -4.0+\rowad) { \includegraphics[width=0.106\textwidth,angle=0]{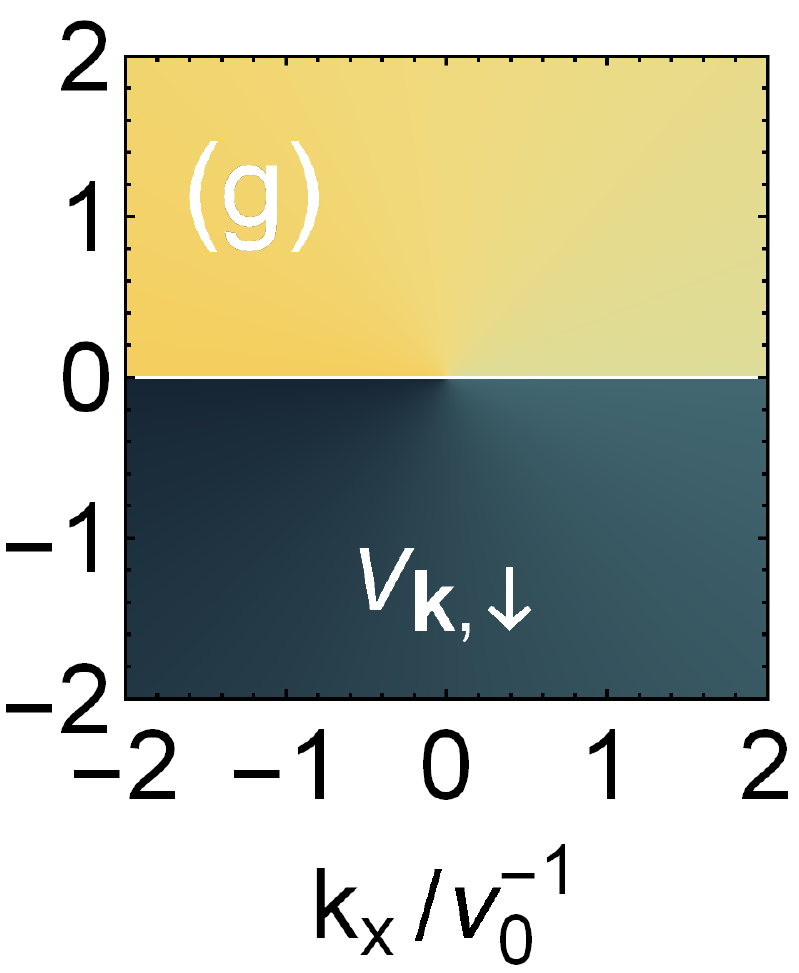}};   
    
     \node[] at (0-2.8, \rowtwo) { \includegraphics[width=0.13\textwidth,angle=0]{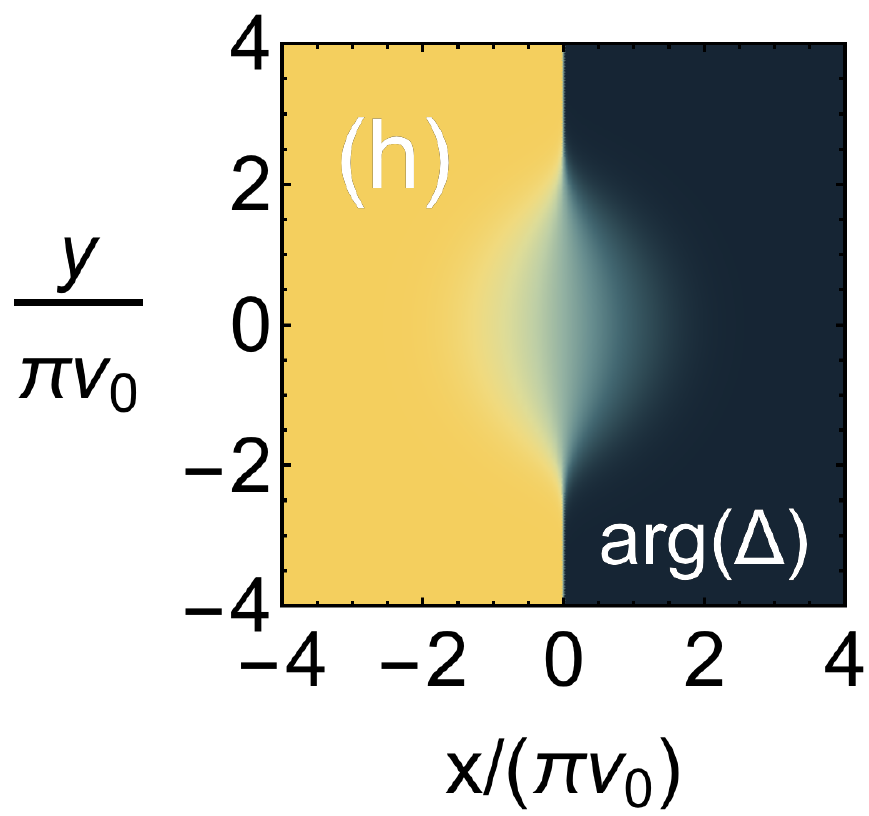}};   
  \node[] at (1.9-2.5,\rowtwo) { \includegraphics[width=0.106\textwidth,angle=0]{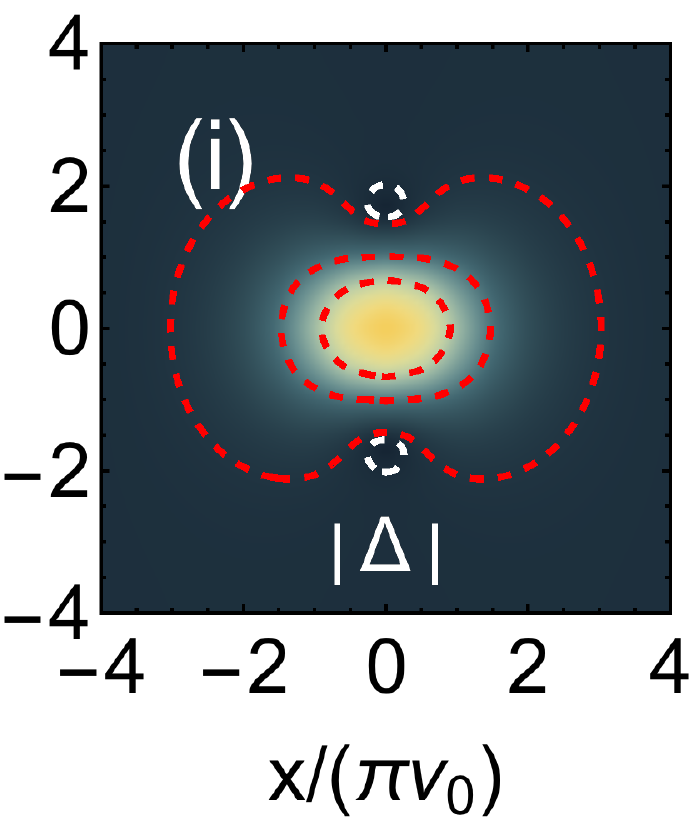}};       
     \node[] at (3.8-2.5, \rowtwo) { \includegraphics[width=0.106\textwidth,angle=0]{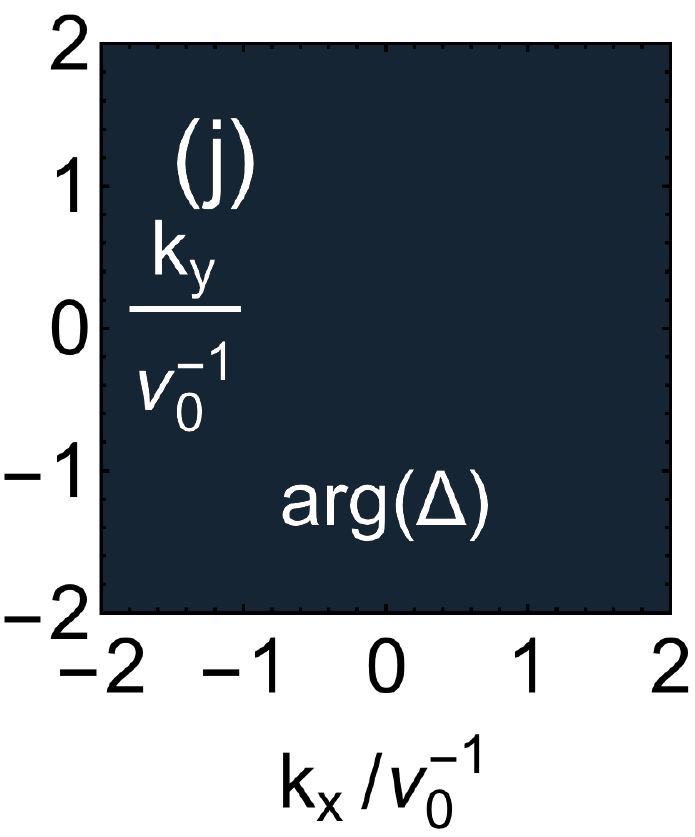}};   
    \node[] at (5.7-2.5, \rowtwo) { \includegraphics[width=0.106\textwidth,angle=0]{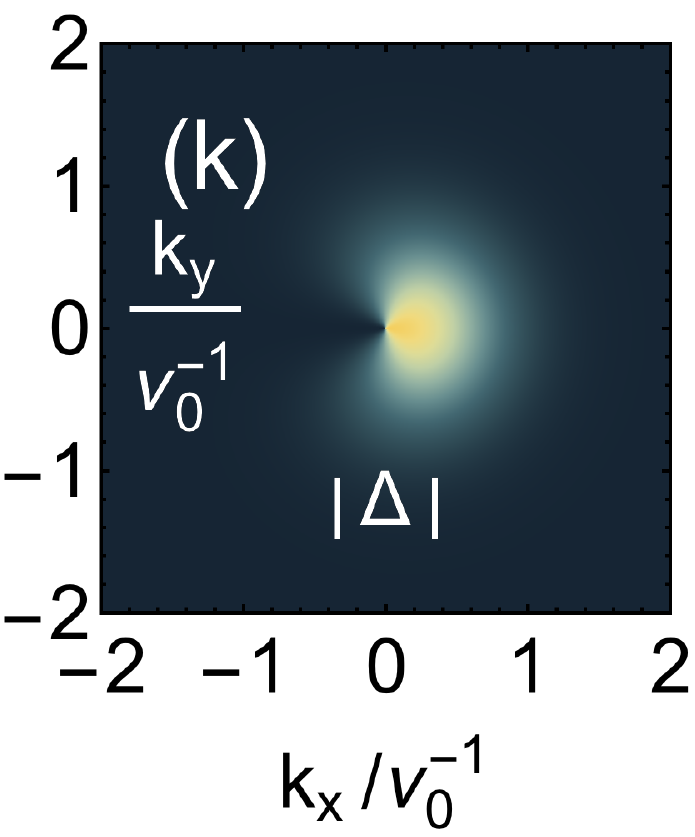}};   

\end{tikzpicture}
      \caption{\label{fig:2} The initial condition at $t = 0$ for the band $\eta_0$. Here $\kappa = \{ u, v\}$, $\nu = \nu_0$, and $g = -1$. (a) Combined spin-$\ua$ and spin-$\da$ density, given by $\nu^2 \rme^{-2\nu^2 k^2}/(\nu_0^2 \pi^3)$. (b, c) Spin-$\ua,\da$ density, given by $\nu^2 \left( \pm k_\x + k \right)\rme^{-2\nu^2 k^2}/(2\nu_0^2  k \pi^3)$ respectively. The colouring and axes are the same as in (a). (d-g) Corresponding phase profiles. (h-k) The superfluid gap, which contains a vortex dipole in real space [Eq.~\eqref{eqn:gapt0realanalytic}]; in momentum space $\Delta(\k,0) = -2 g \left|\kappa_{\k,\ua} \right|^2 \in \mathbb{R}$. The colouring of (a) is scaled by a factor of 2 for (k). In (i) we show four contours the outermost of which is localised at the vortex cores (white).}
\end{figure}

Let us consider an initial vortex dipole-like state in the gap as shown in Fig.~\ref{fig:2}. In the helicity basis, the initial condition is $\phi^\pm_\r(0) = \begin{pmatrix}
\phi^\pm_0(\r) &  \chi^\pm_0(\r) \end{pmatrix}^\mathrm{T}$, where $\phi^\pm_0(\r)  = \frac{\nu_0}{\sqrt{2\pi}\nu} \exp{\left(- \frac{r^2}{4 \nu^2} \right)} = \mathcal{F}_\r^{-1}\left[\phi_0^\pm(\k)\right] = \frac{\nu \nu_0}{\pi\sqrt{2\pi}}\int \d^2 k \exp{\left( -\nu^2 k^2) - \rmi \k\cdot \r\right)}$ is a Gaussian, and $\mathcal{F}_\r^{-1}$ is the inverse Fourier transformation evaluated at $\r$. We take $\chi^\pm_0(\r) = \phi^\pm_0(\r)$ (as in Fig.~\ref{fig:2}). Then in momentum space, $|u^\pm_{\k,\eta_0}(t)|^2 + |v^\pm_{\k,\eta_0}(t)|^2 = 2\phi^\pm_0(\k) = 2\chi^\pm_0(\k) = \nu^2\rme^{-2\nu^2 k^2}/(\nu_0^2 \pi^3)$ is  a constant of motion because the time-evolution operator $\rme^{-\frac{\rmi}{\hbar} H^\mathrm{H}_\pm t}$ is unitary. In momentum space (Fig.~\ref{fig:2}(j,k)), the gap is real-valued but skewed towards positive $k_\x$ momenta. In real space (Fig.~\ref{fig:2}(h,i)), the gap contains a vortex dipole flow field. The key benefit of the helicity basis is that not only is the Hamiltonian decoupled, but the vortices are transformed away into the definition of the basis itself allowing us to work in terms of Gaussians.


\textit{Discussion}
The Bogoliubov quasi-particles are quantum superpositions of the underlying fermionic particles and their holes; in a measurement $\left|u_\r^\pm \right|^2$ ($\left|v_\r^\pm \right|^2$) corresponds to the probability that the excitation is a hole (particle) of helicity $\pm$. More interesting, however, is the superfluid gap parameter $\Delta(\k,t)$, which is evaluated using Eq.~\eqref{eqn:GapEqHelicityBasis}. With the Gaussian initial condition, the number density equation~\eqref{eqn:NumEqHelicityBasis} evaluates to a Gaussian profile, but in the gap we obtain two vortices [Fig.~\ref{fig:2}h, and Eq.~\eqref{eqn:gapt0realanalytic}].

In fact, the procedure is more general. To generate the real-space vortex solutions we require effective $p$-wave pairing, which for Eq.~\eqref{eqn:BdGSolnuv-momspace} was obtained in the helicity frame. However, $p$-wave pairing can emerge in an $s$-wave system also due to Rashba spin-orbit coupling~\cite{Zhang2008}. In this case, the one-body Hamiltonian can be written as~\cite{Zhang2008}
\begin{equation}
\label{eqn:matH2x2pm-dasSarma}
K_0 = \begin{pmatrix}
\epsilon_\kk - \bar{\mu} + h & -\lambda \left(k_\y - \rmi k_\x \right) \\
-\lambda \left(k_\y + \rmi k_\x \right)& \epsilon_\kk - \bar{\mu} - h
\end{pmatrix}.
\end{equation}
Using the same Gaussian initial state as before ($ \chi^\pm_0(\k) = 0$), for an exact vortex solution we must take $h = 0$. We find for the Bogoliubov modes of $K_0$ (Appendix~\ref{app:dSsol})
\begin{subequations}
\begin{align}
q_\r(t) &= D_1(r^2,t),\\
w_\r(t) 
 &=     \left(y +\rmi x\right)  D_2(r^2,t),
\end{align}
\end{subequations}
where the function $D_2(r^2,t)$ is given in Eq.~\eqref{eqn:functionDr2}, and $D_1(r^2,t)$ is given in Eq.~\eqref{eqn:functionD2r2}. A prefactor, $y +\rmi x$, characterising a singly-quantised vortex, appears for $w_\r(t)$.

\begin{figure}[t]
\begin{tikzpicture}
\pgfmathsetmacro{\rowad}{0.0}
\pgfmathsetmacro{\rowinc}{-3.0}

 \node[] at (0, 0) { \includegraphics[width=0.18\textwidth,angle=0]{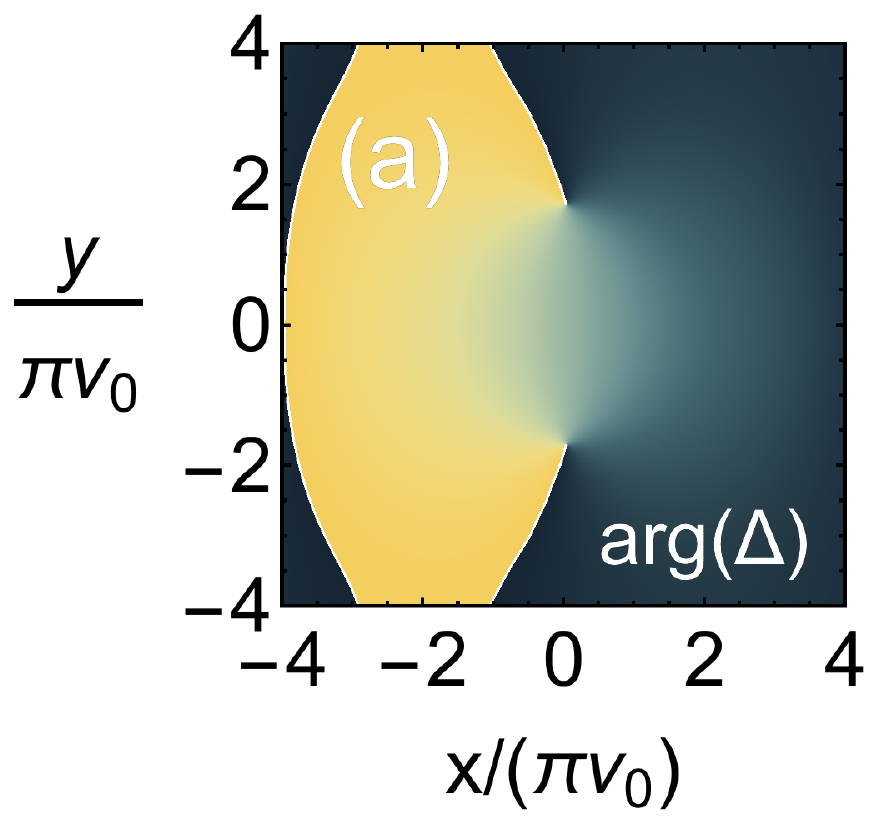}};    
  \node[] at (2.95,0) { \includegraphics[width=0.14\textwidth,angle=0]{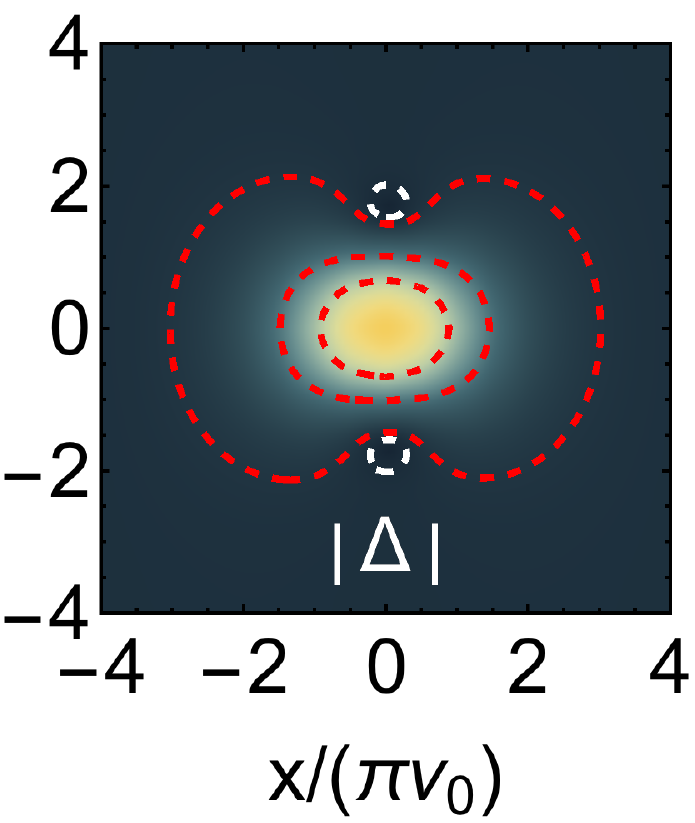}};   
    \node[] at (5.5,0) { \includegraphics[width=0.14\textwidth,angle=0]{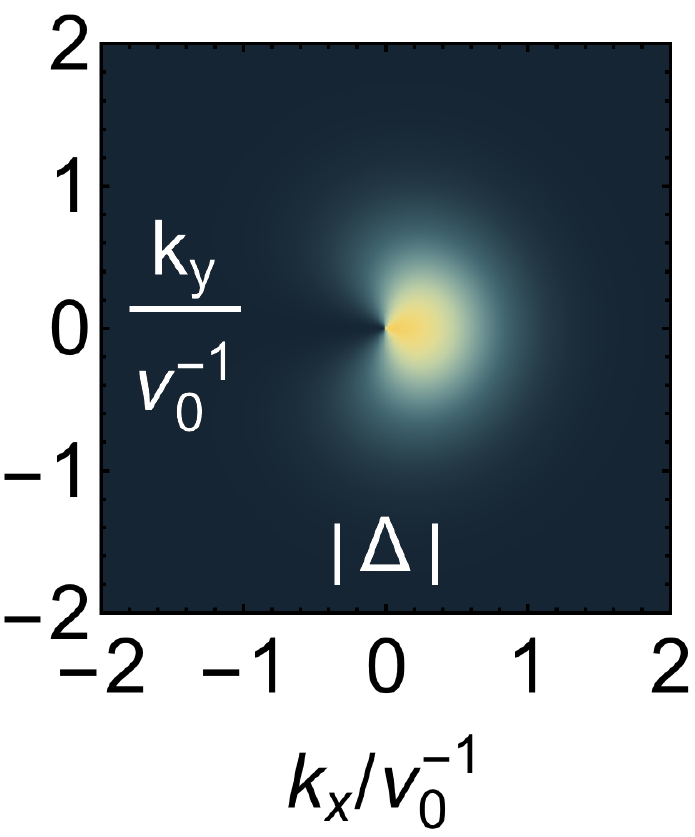}}; 
   \node[] at (0.5, 1.6) { \includegraphics[width=0.1\textwidth,angle=0]{2egl}}; 
   \node[] at (2.95+0.2, 1.6) { \includegraphics[width=0.10\textwidth,angle=0]{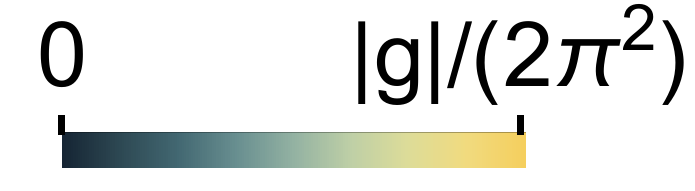}};  
   \node[] at (5.5+0.2, 1.6) { \includegraphics[width=0.10\textwidth,angle=0]{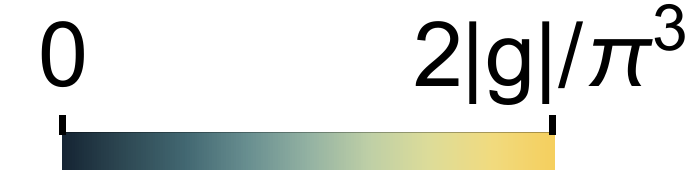}};  
    
 \node[] at (0, \rowinc) { \includegraphics[width=0.18\textwidth,angle=0]{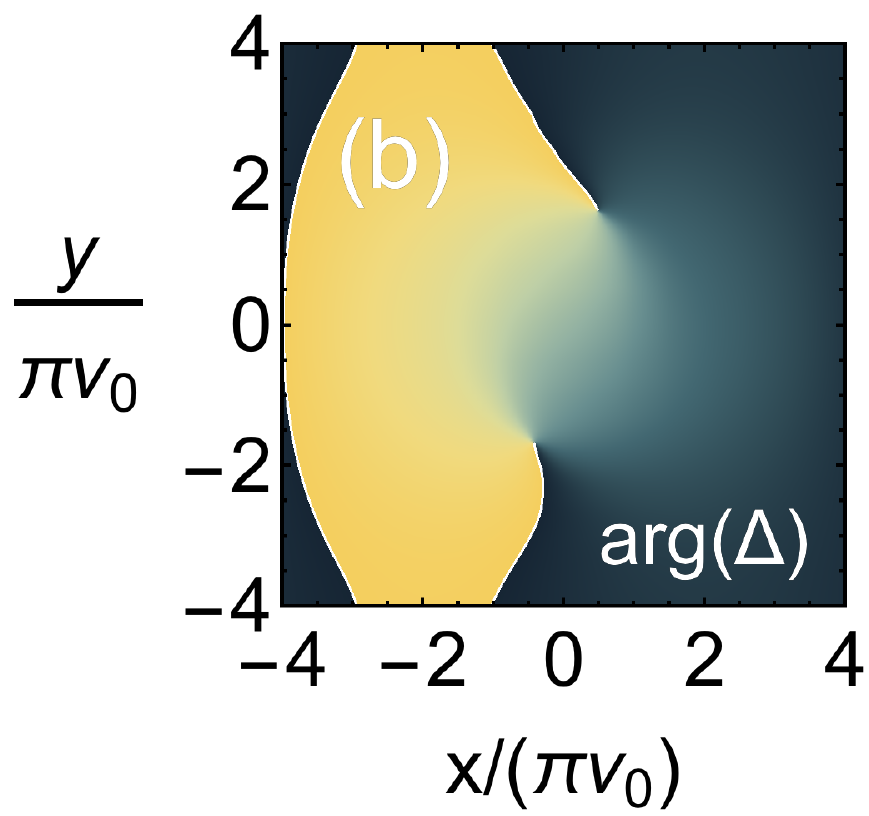}};   
  \node[] at (2.95,\rowinc) { \includegraphics[width=0.14\textwidth,angle=0]{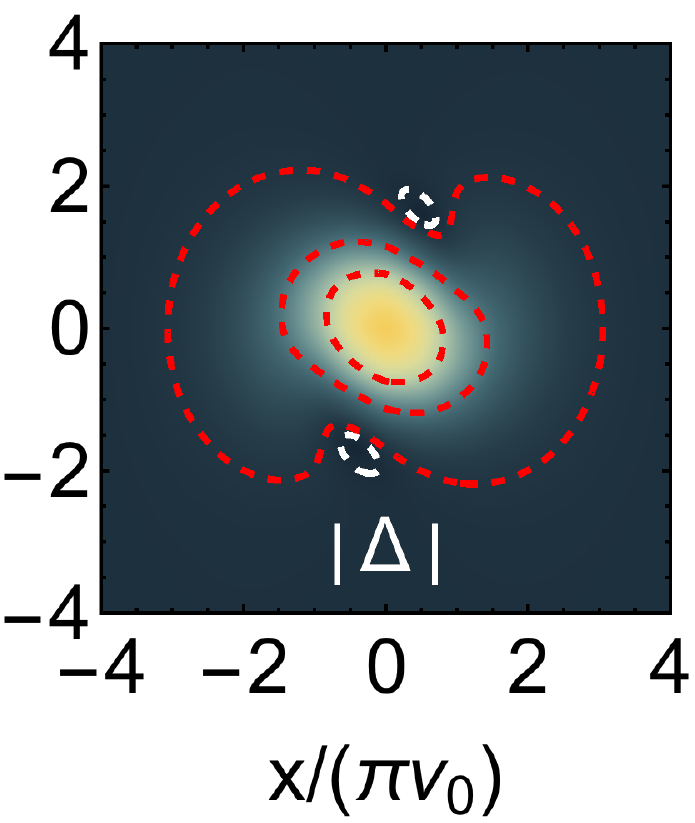}};   
    \node[] at (5.5,\rowinc) { \includegraphics[width=0.14\textwidth,angle=0]{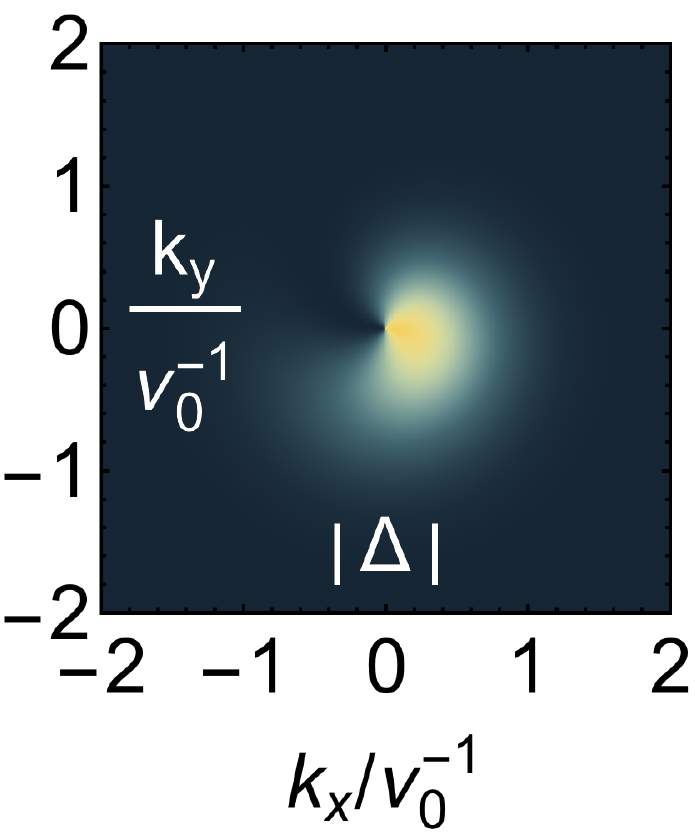}};  
 
  \node[] at (0, 2*\rowinc) { \includegraphics[width=0.18\textwidth,angle=0]{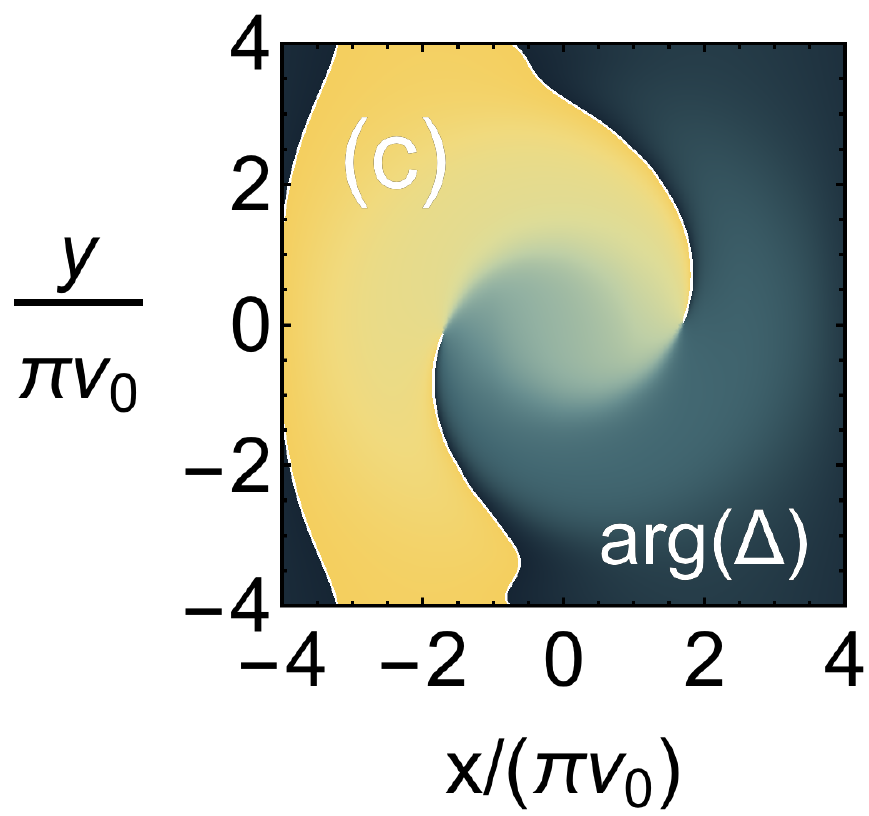}};   
  \node[] at (2.95,2*\rowinc) { \includegraphics[width=0.14\textwidth,angle=0]{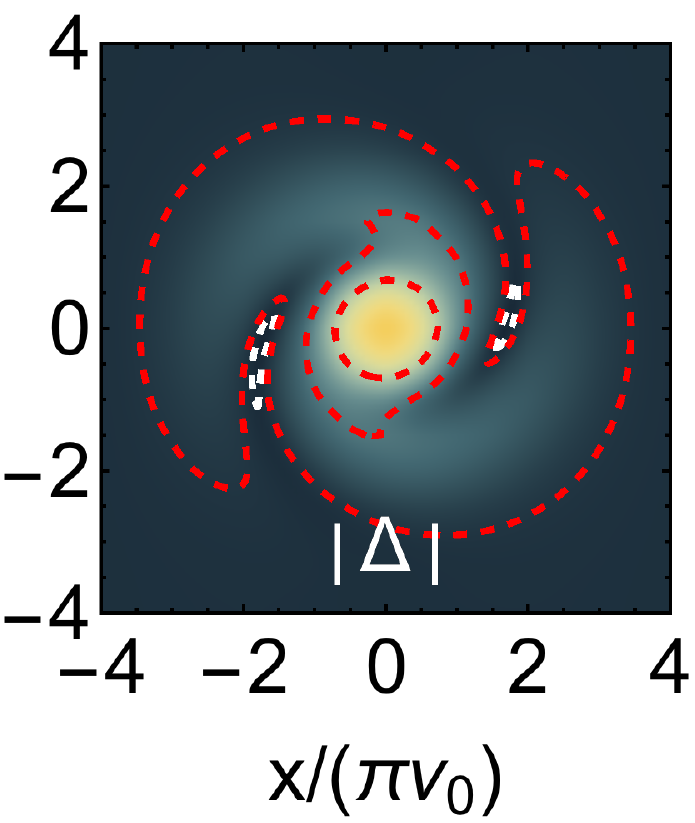}};   
    \node[] at (5.5,2*\rowinc) { \includegraphics[width=0.14\textwidth,angle=0]{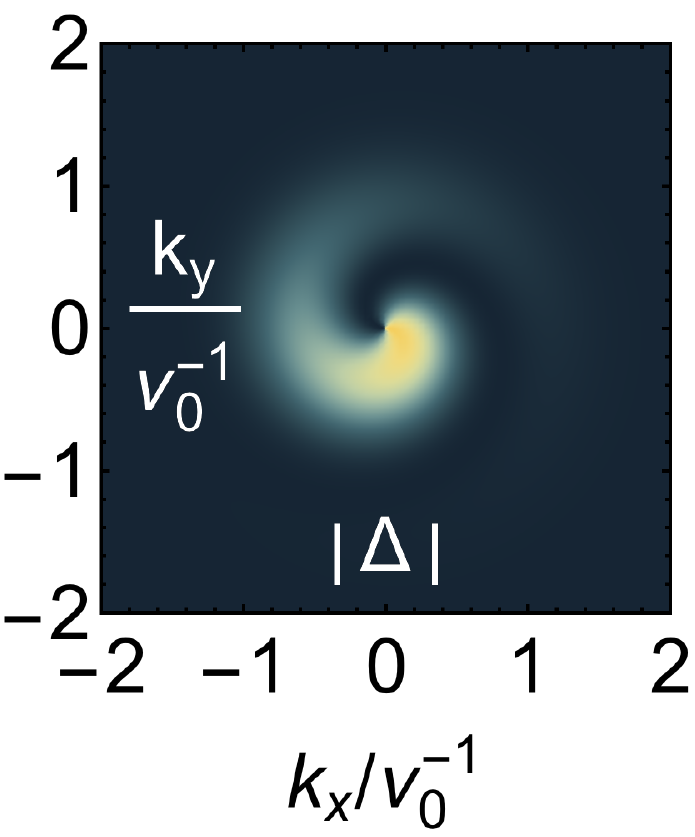}};  
 
  \node[] at (0, 3*\rowinc) { \includegraphics[width=0.18\textwidth,angle=0]{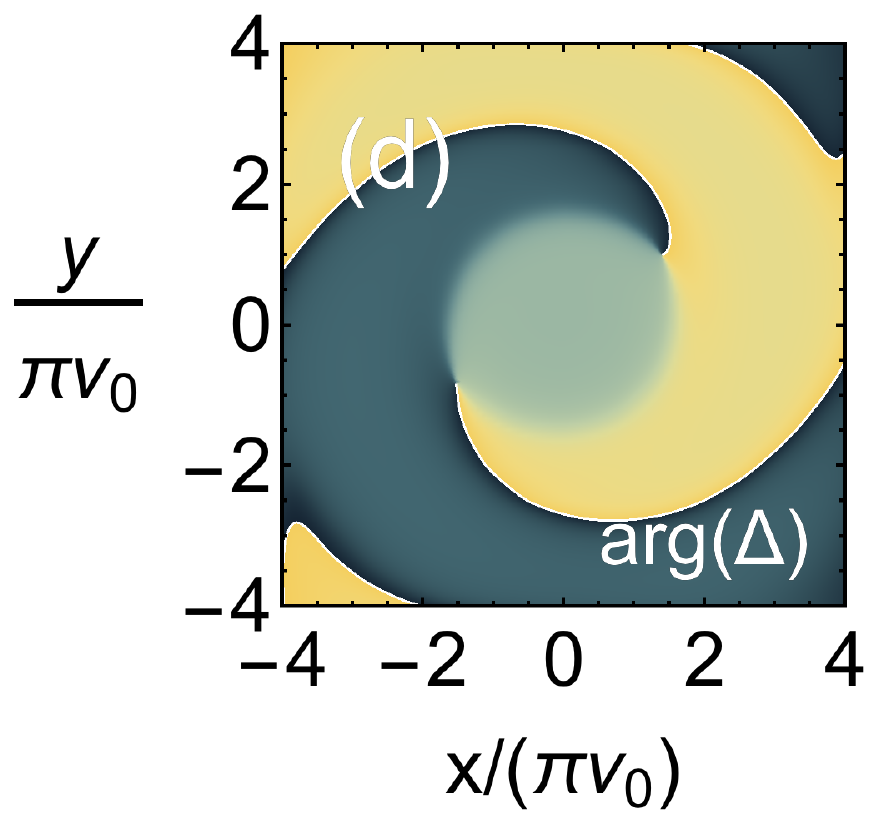}};   
  \node[] at (2.95,3*\rowinc) { \includegraphics[width=0.14\textwidth,angle=0]{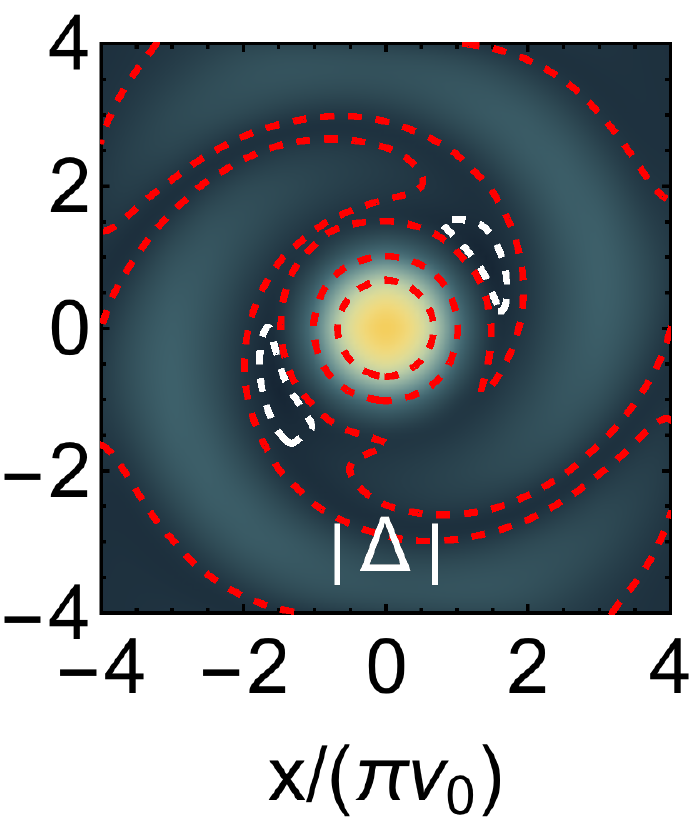}};   
    \node[] at (5.5,3*\rowinc) { \includegraphics[width=0.14\textwidth,angle=0]{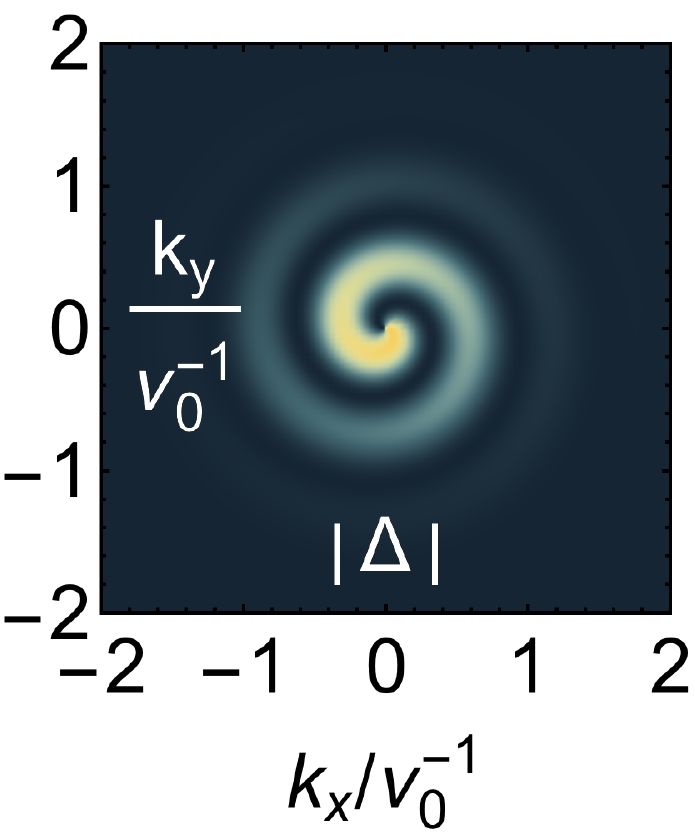}};

\end{tikzpicture}
      \caption{\label{fig:4} The superfluid order parameter $\Delta$ at $t|g|/\hbar = 0.1$ as a function of SO coupling (here $\nu_0 = \nu$). The real-space vortex dipole precesses as a result of the Rashba SO coupling $\lambda$. In momentum space $\Delta(\k,t) \in \mathbb{R}$. The contours are the same as in Fig.~\ref{fig:2}i, with the vortex cores in white. Here $\lambda \nu_0 / |g| $ is (a) 0.0, (b) 5.0, (c) 20.0, and (d) 50.0. Changing $\lambda \to - \lambda$ equals to reflection of all three colums about the $x$-axis, and the precession direction is reversed.}
\end{figure}

Using the solution~\eqref{eqn:BdGSolnuv-momspace}, one must solve the self-consistency conditions~\eqref{eqn:SCHelicityBasis} at every moment in time. Of course, the gap equation need not be satisfied if in the case of a superconductor the gap is induced through the proximity effect such that the order parameter is simply inherited from an adjacent $s$-wave superconductor~\cite{1367-2630-13-6-065004}. The simplest possible case is the approximation of a fully coarse-grained effectively homogeneous system, which we discuss in Appendix~\ref{sec:appselfc}. Then, a self-consistent solution can be found such that the order parameter decays as a function of time (Fig.~\ref{fig:selfcons}); we call this solution a `transient Fermi superfluid'. While considering the more flexible $\k$-dependence in the gap and chemical potential can further lower the free energy resulting in more accurate a solution, the self-consistency equations become significantly more complicated to solve.

We take the SO coupling to be a given constant over the entire system. As a simplification, we also take the chemical potential to be a constant. Strictly speaking this introduces an approximation, but we find numerically that the gap remains unchanged for a wide range of the chemical potential between $t = 0$ and $t|g|/\hbar = 0.1$ ($\mu$ is not adjusted for the initial condition). When evaluating the gap at $t|g|/\hbar = 0.1$, we use the exact value at $t = 0$ shown in Fig.~\ref{fig:2}. SO coupling makes the vortex dipole in the gap precess (anti-)clockwise for $\lambda \nu_0 / |g| > 0 $ ($< 0$) (Figs.~\ref{fig:4} and~\ref{fig:6}) at a frequency that increases as the magnitude of $\lambda$ increases. It is computationally demanding to accurately propagate the gap, so we consider only the small time $t|g|/\hbar = 0.1$, and use the gap at $t = 0$ (Fig.~\ref{fig:2}) for the numerical evaluation of the gap at this time. Moreover, a larger SO coupling strength increases the time-dependence of the gap, which we have ignored when constructing the solution~\eqref{eqn:BdGSolnuv-momspace}. Therefore, this is not an exact solution for the full spatiotemporal self-consistency conditions, but nevertheless captures the main physical effect and qualitative aspects of the vortex dipole precession due to Rashba SO coupling.

\textit{Conclusions}
We have theoretically predicted that a vortex dipole in the superfluid order parameter of a spin-balanced $s$-wave Fermi supefluid with Rasbha spin-orbit coupling will precess if the SO coupling is non-zero. In the approximation of a homogeneous gap, we showed that the superfluid order decays over time thus corresponding to a transient Fermi superfluid. In addition to demonstrating time dynamics of the vortex dipole due to SO coupling alone, our general solution forms a starting point for perturbative studies of vortex dynamics under the effects of broken time-reversal symmetry, that is, population imbalance ($h \neq 0$) and the topological phase corresponding to $h > \sqrt{\bar{\mu}^2+\absD^2}$, which further requires spin-orbit coupling. In the topological phase, vortex precession in principle corresponds to braiding of Majorana fermions, which is a key element of topological quantum computation. Studying the vortex dipole precession effect in a topological Fermi gas will build towards the important goal of experimentally controlling and manipulating quantum information stored in Majorana-based qubit systems.

\acknowledgements
This work was supported by the Institute for Basic Science in Korea (IBS-R024-D1). I would like to thank Yuri Rubo for fruitful discussions.

\bibliographystyle{apsrev4-1}

\bibliography{Helicity_TF}

\onecolumngrid
\appendix
\section{\label{sec:appselfc}Self-consistency equations}
The microscopic Hamiltonian of the Fermi gas reads
\begin{equation}
\label{eqn:micHmom}
\begin{split}
H - \sum_\sigma \mu_\sigma N_\sigma &= \frac{1}{\Omega}\sum_{\k\sigma} \left(\epsilon_\k - \mu_\sigma\right)c_{\k,\sigma}^\dagger c_{\k,\sigma} +\frac{1}{\Omega} \sum_\k \left(\lambda k \rme^{\rmi \varphi_\k}c_{\k,\ua}^\dagger c_{\k,\da}  + \text{h.c.} \right) + \frac{1}{\Omega^2} \sum_{\k,\kp} V_{\k,\kp} c_{\k,\ua}^\dagger c_{-\k,\da}^\dagger c_{-\kp,\da} c_{\kp,\ua}, 
\end{split}
\end{equation}
where $N_\sigma$ is the total number of atoms with spin $\sigma = \left\lbrace \ua,\da \right\rbrace$. For singlet pairing $V_{\k,\kp} = -V_{\k,-\kp}$. Under the assumption that Cooper pairing represents the actual physics, the order parameter (relative wavefunction of the Cooper pairs) is defined as $\Delta_\k \equiv - \Omega^{-1} \sum_{\kp} V_{\k,\kp} \langle c_{-\kp \da} c_{\kp \ua} \rangle =  \Omega^{-1} \sum_{\kp} V_{\k,\kp} \langle  c_{\kp \ua} c_{-\kp \da}\rangle$. If we now make the \textit{mean-field approximation} of ignoring all the other correlations apart from the pair correlations captured by the order parameter, and focus on generalised attractive $s$-wave pairing where $V^\mathrm{s-wave}_{\k,\kp} = g \delta_{\k,\kp}< 0$, we obtain the mean-field Hamiltonian and the starting point of our work, Eq.~\eqref{eqn:HswaveFermiSF-p}. The interaction term in Eq.~\eqref{eqn:micHmom} becomes
\begin{equation}
\label{eqn:micHmom-meanfield}
+ \frac{1}{\Omega^2} \sum_{\k,\kp} V_{\k,\kp} c_{\k,\ua}^\dagger c_{-\k,\da}^\dagger c_{-\kp,\da} c_{\kp,\ua} \to - \frac{1}{\Omega} \sum_{\k}\left( \Delta_\k^* c_{-\k,\da} c_{\k,\ua} + \Delta_\k c_{\k,\ua}^\dagger c_{-\k,\da}^\dagger\right).
\end{equation}
Explicity, the mean-field Hamiltonian matrix reads
\begin{equation}
\label{eqn:micHmom1}
H - \sum_\sigma \mu_\sigma N_\sigma =  \frac{1}{\Omega}\sum_\kk \, \begin{pmatrix} c_{\k,\ua}^\dagger & c_{\k,\da}^\dagger & c_{-\k,\ua} & c_{-\k,\da} \end{pmatrix} 
\begin{pmatrix}
\epsilon_\k - \mu_\ua & \lambda k \rme^{\rmi \varphi_\k} & 0 &-\Delta(\k,t) \\
\lambda k \rme^{-\rmi \varphi_\k} & \epsilon_\k - \mu_\da & \Delta(\k,t) &0 \\
 0& \Delta^*(\k,t) & -\left( \epsilon_\k - \mu_\ua \right) & \lambda k \rme^{-\rmi \varphi_\k}  \\
-\Delta^*(\k,t) & 0 & \lambda k \rme^{\rmi \varphi_\k} & -\left( \epsilon_\k - \mu_\da \right)
\end{pmatrix}
\begin{pmatrix}c_{\k,\ua} \\ 
c_{\k,\da} \\
 c_{-\k,\ua}^\dagger \\
  c_{-\k,\da}^\dagger \end{pmatrix},
\end{equation}
which is Eq.~\eqref{eqn:HswaveFermiSF-p} in the main text.

Let us now denote the eigenvectors of $H(\k)$ by the Bogoliubov $u,v$ notation; we define the eigenvector $\begin{pmatrix}
u_{\k\eta,\ua} &
u_{\k\eta,\da}& v_{-\k\eta,\ua}& v_{-\k\eta,\da}\end{pmatrix}^\mathrm{T}$ to have eigenvalue $E_\eta$. Then $\begin{pmatrix}
v_{\k\eta,\ua}^* &
v_{\k\eta,\da}^* & u_{-\k\eta,\ua}^*& u_{-\k\eta,\da}^*\end{pmatrix}^\mathrm{T}$ is also an eigenvector with energy $-E_\eta$, which follows from the particle-hole symmetry of $H(\k)$. The energies $E_\eta$ can be easily obtained by diagonalising $H(\k)$, which are also reported in Refs.~\cite{Jiang2011,Zhou2011}; we find $-E_2 = E_3 = \left|E_{\k,-}\right|$ and $-E_1 = E_4 = \left|E_{\k,+}\right|$, where $E_{\k,\pm} = \sqrt{\left(\epsilon_\k -\bar{\mu}\right)^2 + \absD^2 + h^2 + \lambda^2 k^2 \pm 2\sqrt{\left(h^2+\lambda^2 k^2 \right)\left(\epsilon_\k -\bar{\mu}\right)^2 + h^2\absD^2}}$.

At zero temperature, in terms of the Bogoliubov $u,v$ notation, the order parameter equation becomes
\begin{equation}
\label{eqn:self_cons_orderp-p}
\begin{split}
\Delta_\k = -\frac{1}{\Omega}\sum_{\kp,\eta} V_{\k,\kp} \langle  c_{\kp, \uparrow} c_{-\kp, \downarrow} \rangle 
&\stackrel{T = 0}{=}\frac{g}{\Omega}\left( \sum_{E_\eta<0}  u_{\k\eta,\ua} v_{-\k\eta,\da}^* + \sum_{E_\eta>0}  u_{-\k\eta,\da} v_{\k\eta,\ua}^* \right)\\
&=\frac{2g}{\Omega}\sum_{E_\eta<0}  u_{\k\eta,\ua} v_{-\k\eta,\da}^*,
\end{split}
\end{equation}
which simply sums over the particle-hole symmetry $\left(u_{\k,\ua} v_{-\k,\da}^*\right)_{E_{2(1)}} = \left(u_{-\k,\da} v_{\k,\ua}^*\right)_{E_{3(4)}}$.

We now use the `helicity transformation'~\eqref{eqn:HelBasisTF} to write
\begin{equation}
\label{eqn:HelBasisTF1}
\begin{pmatrix}
u_{\k,\ua}\\
u_{\k,\da}
\end{pmatrix}
= \frac{1}{\sqrt{2}}
\begin{pmatrix}
1 & \rme^{\rmi \varphi_\k}\\
\rme^{-\rmi \varphi_\k} & -1
\end{pmatrix}
\begin{pmatrix}
u_\k^+\\
u_\k^-
\end{pmatrix},
\qquad
\begin{pmatrix}
v_{-\k,\ua}\\
v_{-\k,\da}
\end{pmatrix}
= \frac{1}{\sqrt{2}}
\begin{pmatrix}
1 & -\rme^{-\rmi \varphi_\k}\\
-\rme^{\rmi \varphi_\k} & -1
\end{pmatrix}
\begin{pmatrix}
v_{-\k}^+\\
v_{-\k}^-
\end{pmatrix}.
\end{equation}
By direct substitution, the order parameter equation~\eqref{eqn:self_cons_orderp-p} becomes
\begin{equation}
\label{eqn:self_cons_orderp-hel-p}
\begin{split}
\Delta &=\frac{2g}{\Omega} \sum_{E_\eta<0}   u_{\k\eta,\uparrow} v_{-\k\eta,\downarrow}^*\\
&=\frac{g}{\Omega} \sum_{E_\eta<0}   \left(u_{\k\eta}^+ +\rme^{\rmi \varphi_\k} u_{\k\eta}^- \right) \left(-\rme^{-\rmi \varphi_\k}\bar{v}_{-\k\eta}^+ - \bar{v}_{-\k\eta}^- \right)\\
&=-\frac{g}{\Omega} \sum_{E_\eta<0} \left( 
\rme^{-\rmi \varphi_\k} u_{\k\eta}^+ \bar{v}_{-\k\eta}^+
+
u_{\k\eta}^+ \bar{v}_{-\k\eta}^-
+
u_{\k\eta}^- \bar{v}_{-\k\eta}^+
+
\rme^{\rmi \varphi_\k} u_{\k\eta}^- \bar{v}_{-\k\eta}^-
\right),
\end{split}
\end{equation}
where the bar denotes complex conjugation. This is Eq.~\eqref{eqn:GapEqHelicityBasis} in the main text.

Similarly, to determine the chemical potential, the number density 
\begin{equation}
n_\sigma(\k,t) = \langle c^\dagger_{\k,\sigma} c_{\k,\sigma}\rangle  =\sum_{E_{\eta}<0} \left|
u_{\k\eta,\sigma} \right|^2 + \sum_{E_{\eta}>0} \left|
v_{\k\eta,\sigma} \right|^2 
\end{equation}
for spin $\sigma$ fermions must be solved self-consistently. In the helicity basis [Eq.~\eqref{eqn:HelBasisTF1}], we have
\begin{equation}
\begin{split}
n =\frac{1}{\Omega} \sum_{\k,\sigma} n_\sigma(\k) &= \frac{1}{2\Omega}\sum_{\k,E_{\eta}<0}\left( \left|
u_{\k\eta,\ua} \right|^2 +  \left|
u_{\k\eta,\da} \right|^2\right) + \frac{1}{2\Omega}\sum_{\k,E_{\eta}>0}\left( \left|
v_{\k\eta,\ua} \right|^2 +  \left|
v_{\k\eta,\da} \right|^2\right) \\
&= \frac{1}{4\Omega}\sum_{\k,E_{\eta}<0}\left|
\begin{pmatrix}
1 & \rme^{\rmi \varphi_\k}\\
\rme^{-\rmi \varphi_\k} & -1
\end{pmatrix}
\begin{pmatrix}
u_{\k\eta}^+\\
u_{\k\eta}^-
\end{pmatrix}
\right|^2 
+
\frac{1}{4\Omega}\sum_{\k,E_{\eta}>0}\left|
\begin{pmatrix}
1 & \rme^{-\rmi \varphi_\k}\\
\rme^{\rmi \varphi_\k} & -1
\end{pmatrix}
\begin{pmatrix}
v_{\k\eta}^+\\
v_{\k\eta}^-
\end{pmatrix}
\right|^2
\\
&=\frac{1}{2\Omega}\sum_{\k,E_{\eta}<0}\left( |
u_{\k\eta}^+ |^2 +  |
u_{\k\eta}^- |^2\right)+\frac{1}{2\Omega}\sum_{\k,E_{\eta}>0}\left( |
v_{\k\eta}^+ |^2 +  |
v_{\k\eta}^- |^2\right) .
\end{split}
\end{equation}

\subsection{Coarse-grained case}
In this effectively homogeneous case we sum over $\k$ making $\Delta(\k,t) = \Delta(t)$ and $n(\k,t) = n(t)$, independent of $\k$. When $\chi^\pm_0(\k) = 0$, direct substitution yields the transcendental condition
\begin{equation}
\label{eqn:GapEqHelicityBasis1-1}
\begin{split}
\frac{\Delta}{g} &= 
\beta^2 \sum_{\alpha = \pm}\int_0^\infty \d k \, k \,\rme^{-2\nu^2 k^2} \Delta \left[- \frac{  \rmi  S_\alpha C_\alpha }{\absEa} -   \frac{S_\alpha^2 \xi_\alpha}{\absEa^2}\right],
 \end{split}
\end{equation}
where $\beta \equiv \nu/(2 \pi^2) $, $\xi_\alpha \equiv \epsilon_\k - \bar{\mu} +\alpha \lambda k$, $C_\alpha \equiv \cos{\left(\frac{\absEa t}{\hbar}  \right)}$, 
$S_\alpha \equiv \sin{\left(\frac{\absEa t}{\hbar}  \right)}$, and we used the 2D continuum limit $\Omega^{-1}\sum_\k \to (2\pi)^{-1} \nu_0^2 \int_0^\infty \d k \, k$. The cross terms in Eq.~\eqref{eqn:GapEqHelicityBasis} vanish by symmetry. There cannot be a solution to Eq.~\eqref{eqn:GapEqHelicityBasis1-1} if the imaginary term does not vanish. Equation~\eqref{eqn:GapEqHelicityBasis1-1} must, in principle, be supplemented with the number density equation~\eqref{eqn:NumEqHelicityBasis}. However, the normalised initial condition $\phi_\k^\pm(0)$ adjusts $n = 1/(16 \pi^4)$, independent of the three parameters $\{\lambda, \Delta, \bar{\mu}\}$ and time. The Gaussian initial condition also makes the integral over $k$ convergent, which would require renormalisation of $g$ in the case of a genuinely homogeneous Fermi gas~\footnote{While generally convenient, in the homogeneous case a contact interaction $g$ leads to uncontrolled scattering which renders the gap equation logarithmically divergent. In the presence of spin-orbit coupling, the bare contact interaction $g$ is renormalised following the standard relation in two dimensions~\cite{Zhou2011}, $\frac{1}{g} = -\frac{1}{\Omega} \Sigma_\k \frac{1}{ 2\epsilon_\k + E_\mathrm{b}}$, where $E_\mathrm{b}>0$ is the absolute value of the binding energy~\cite{PhysRevLett.62.981,Petrov2000,Petrov2001,Petrov2002} of the
two-particle bound state when $\lambda = 0$. The quantity $E_\mathrm{b}$ in fact parametrises the BEC-BCS crossover.}. 

\begin{figure}[t]
  \centering
    \includegraphics[width=0.45\textwidth,angle=0]{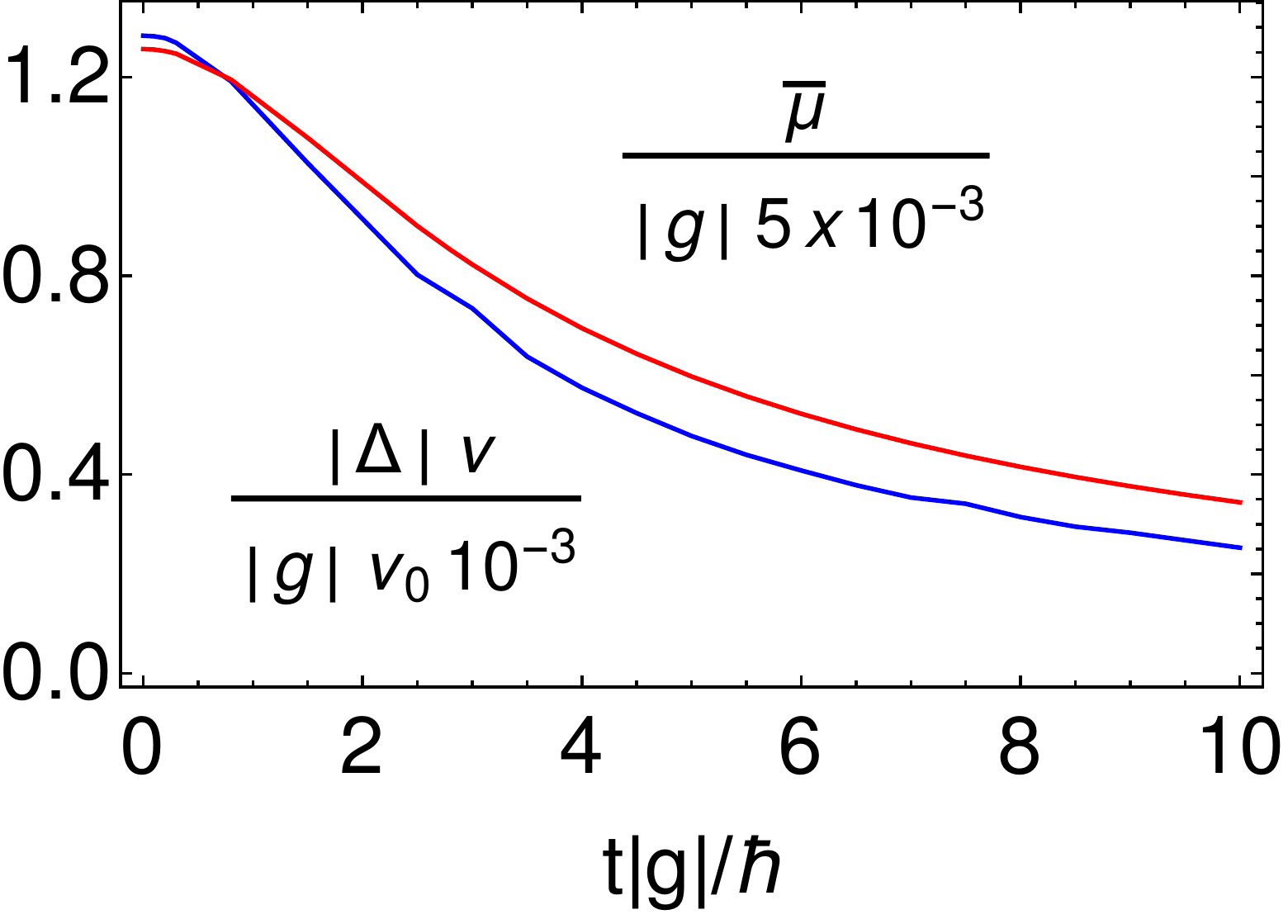}
      \caption{\label{fig:selfcons} Self-consistent values for the homogeneous superfluid order parameter $\absD$ (blue) and chemical potential $\bar{\mu}$ (red). We have set $g < 0$, and constrained spin-orbit coupling to be a constant, $\lambda \nu_0/|g| = 0.5$. The order parameter decays as a function of time, characterising a `transient Fermi superfluid'.}
\end{figure} 

Of all the possible solutions to Eq.~\eqref{eqn:GapEqHelicityBasis1-1}, only the one corresponding to a global minimum of the thermodynamic potential $\Xi = \bra{\Psi}\mathcal{H}\ket{\Psi}$, where $\Psi$ is the BCS wavefunction, is realized in practice. The gap equation then corresponds to the extremum condition $\partial \Xi / \partial \Delta^* = 0$. At $t = 0$, the only possible solution is $\absD = 0$, which can be seen by substitution of the initial condition $\phi_0^\pm(\k)$ into the gap equation. For later times, $\Delta = 0$ remains a trivial solution; in fact, we find that $\absD = 0$ is the only solution that is both (i) a global minimum of $\Xi$; and (ii) a self-consistent solution to the gap equation, and therefore there cannot be any self-consistent superfluid pairing. 

When $\chi^\pm_0(\k) \neq 0$, at $t = 0$ a finite gap is no longer trivially ruled out. Let us now construct the initial condition such that at $t = 0$, when the imprinting of the state $\phi_\r^\pm(0)$ is performed, we have a coarse-grained effectively uniform Fermi superfluid corresponding to the gap $\Delta_0$. Taking $\chi^\pm_0(\k) = \phi^\pm_0(\k)$, direct substitution of the two-component Gaussian initial condition into Eq.~\eqref{eqn:GapEqHelicityBasis}, and integrating over $\k$, yields
\begin{equation}
\label{eqn:Delta0}
\begin{split}
\Delta_0 &=-g \frac{\nu_0}{8\pi^4\nu}.
\end{split}
\end{equation}
As expected, the number density equation evaluates to $n = 1/(8 \pi^4)$, twice as large compared to the case  $\chi^\pm_0(\k) = 0$, and independent of the three parameters $\{\lambda, \Delta, \bar{\mu}\}$ and time. The finite solution $\Delta_0$ will always correspond to the global minimum because $\Xi = \absD^2 + g \frac{\Delta^* \nu_0}{8\pi^4\nu}$ is bounded from below and there are only two stationary points, one at $\Delta = 0$ and one at $\Delta = \Delta_0$. Obtaining a finite gap $\Delta_0$ with the symmetric initial condition means that the emergence of vortices in the hole and particle-like components is also symmetric. The helicity vortices come with opposite circulations [the factors $\rme^{\pm \rmi \varphi_\k}$ in Eq.~\eqref{eqn:BdGSolnuv-u-momspace} and $\rme^{\mp \rmi \varphi_\k}$ in Eq.~\eqref{eqn:BdGSolnuv-v-momspace}], that is, they appear in vortex-antivortex pairs.

\begin{figure}[t]
\begin{tikzpicture}
   
\pgfmathsetmacro{\rowsep}{2.3}
   
 \node[] at (0-2.8, -4.0-0.2) { \includegraphics[width=0.13\textwidth,angle=0]{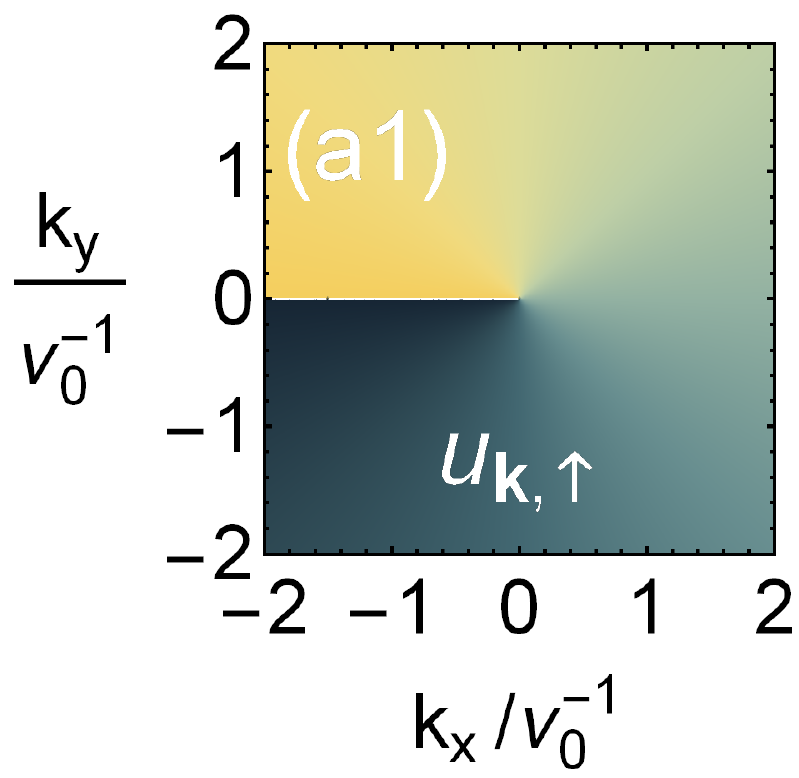}};   
 \node[] at (0-2.5, -2.7-0.2) { \includegraphics[width=0.085\textwidth,angle=0]{2egl}};    
  \node[] at (1.9-2.5, -4.0-0.2) { \includegraphics[width=0.106\textwidth,angle=0]{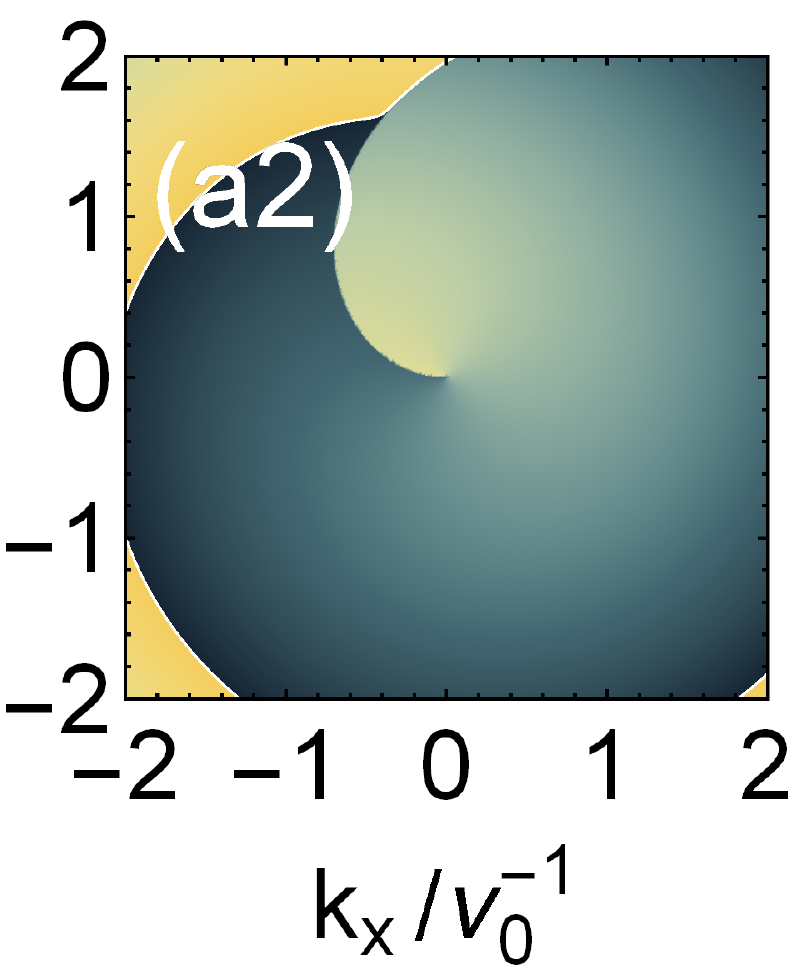}};   
   \node[] at (2.0-2.5, -2.7-0.2) { \includegraphics[width=0.085\textwidth,angle=0]{2egl}}; 
    \node[] at (3.9-2.5, -2.7-0.2) { \includegraphics[width=0.085\textwidth,angle=0]{2egl}};    
   \node[] at (3.8-2.5, -4.0-0.2) { \includegraphics[width=0.106\textwidth,angle=0]{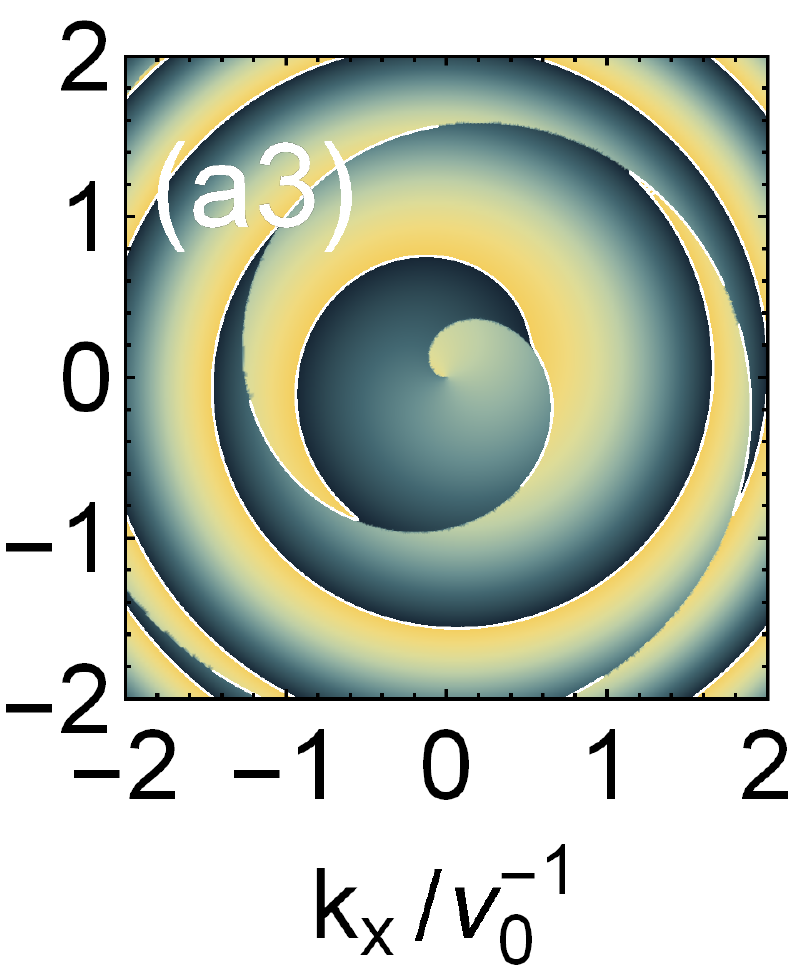}};   
   \node[] at (5.8-2.5, -2.7-0.2) { \includegraphics[width=0.085\textwidth,angle=0]{2egl}};
    \node[] at (5.7-2.5, -4.0-0.2) { \includegraphics[width=0.106\textwidth,angle=0]{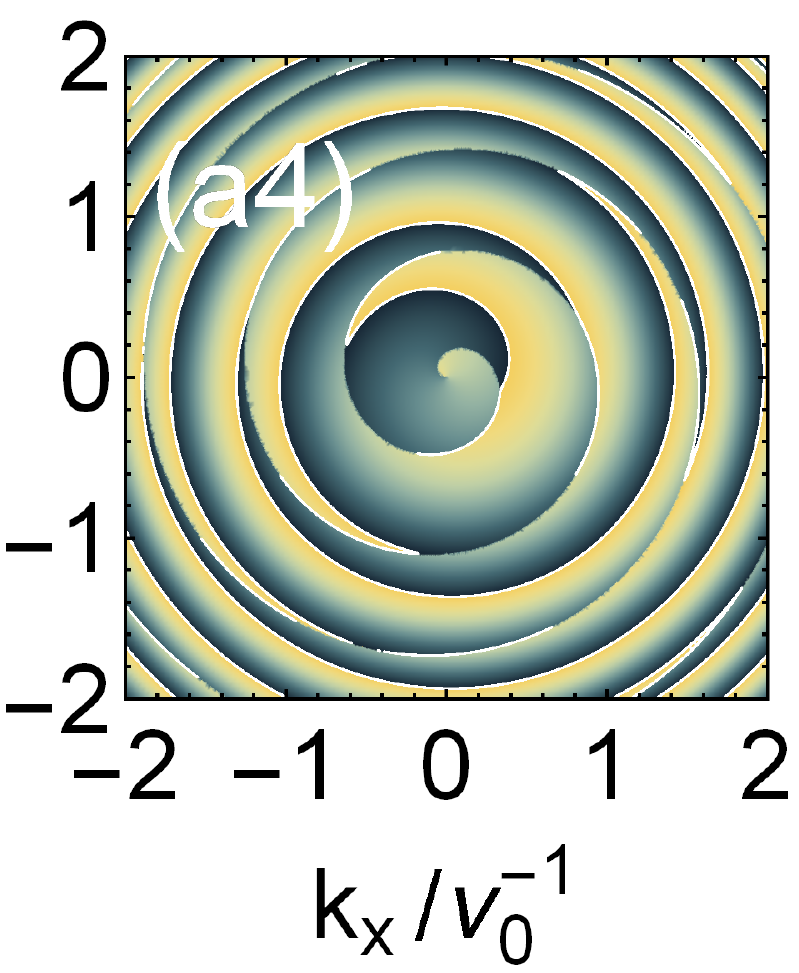}};

 \node[] at (0-2.8, -4.0-0.2-\rowsep) { \includegraphics[width=0.13\textwidth,angle=0]{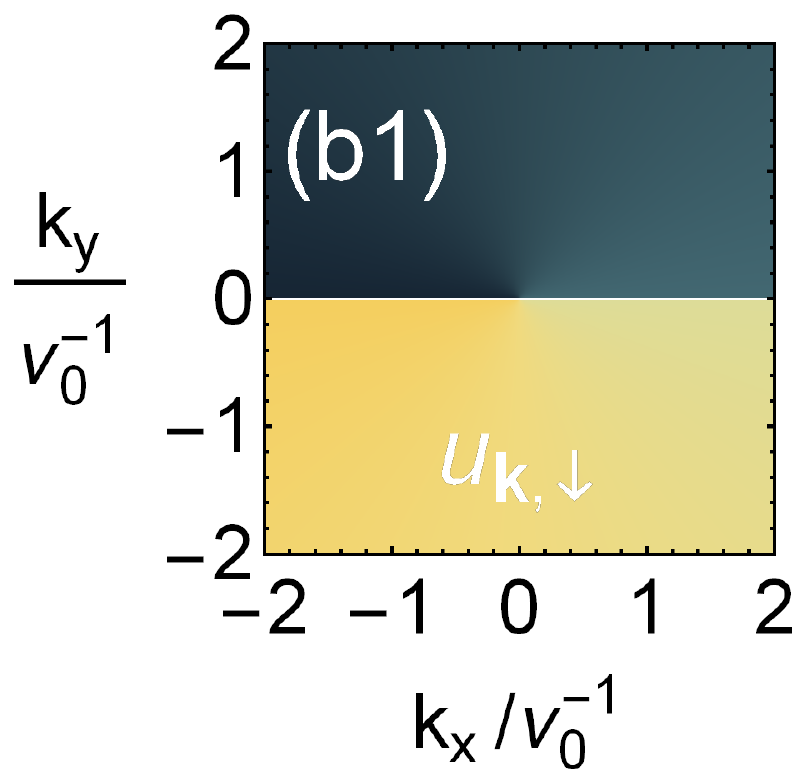}};   
  \node[] at (1.9-2.5, -4.0-0.2-\rowsep) { \includegraphics[width=0.106\textwidth,angle=0]{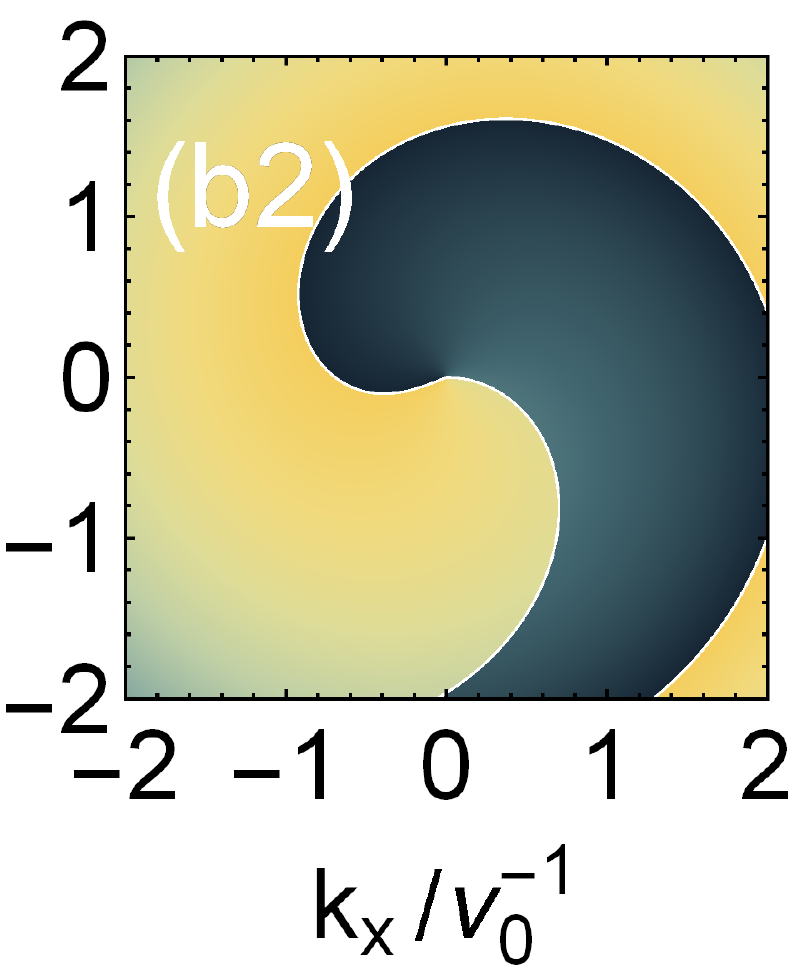}};   
   \node[] at (3.8-2.5, -4.0-0.2-\rowsep) { \includegraphics[width=0.106\textwidth,angle=0]{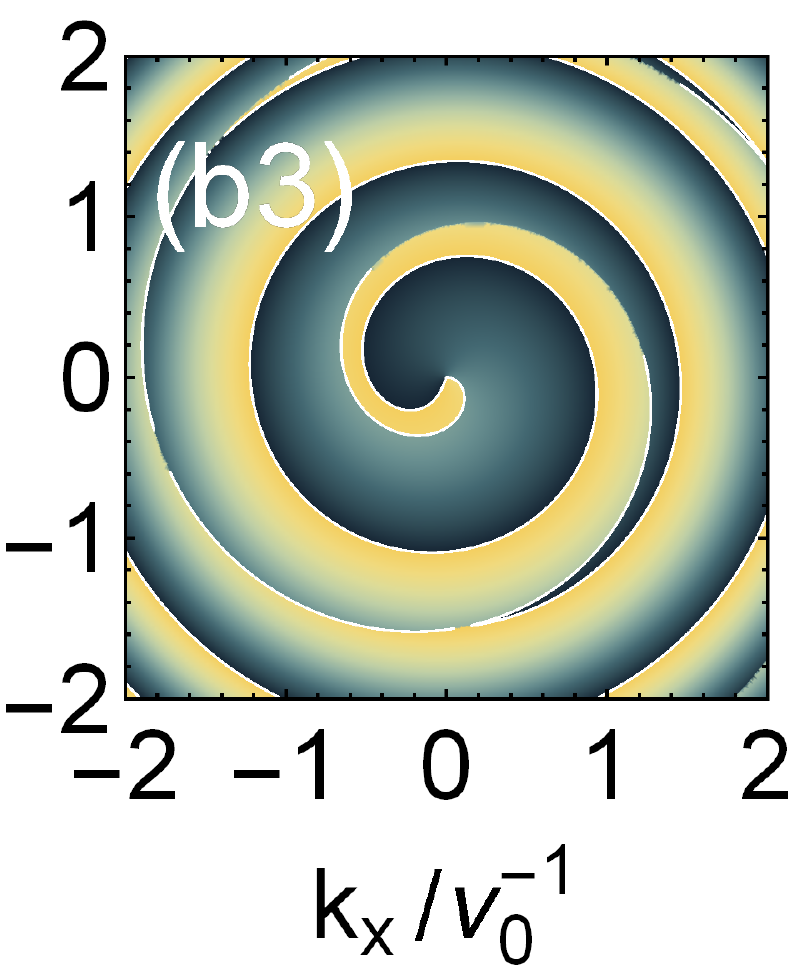}};   
    \node[] at (5.7-2.5, -4.0-0.2-\rowsep) { \includegraphics[width=0.106\textwidth,angle=0]{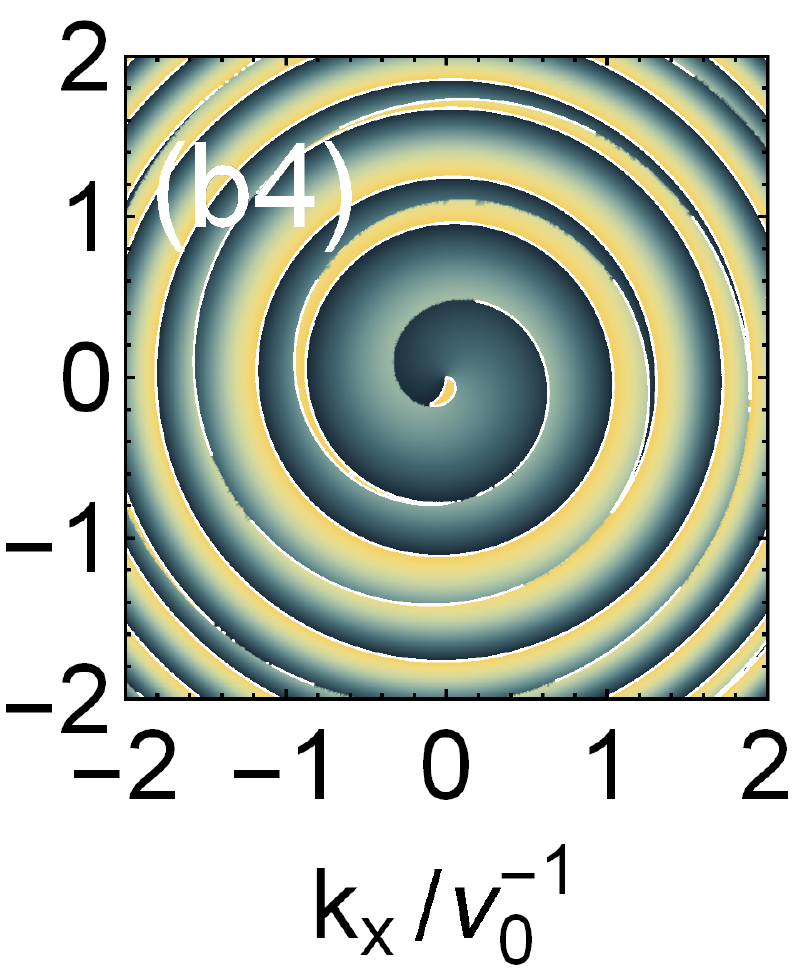}};       
   
 \node[] at (0-2.8, -4.0-0.2-2*\rowsep) { \includegraphics[width=0.13\textwidth,angle=0]{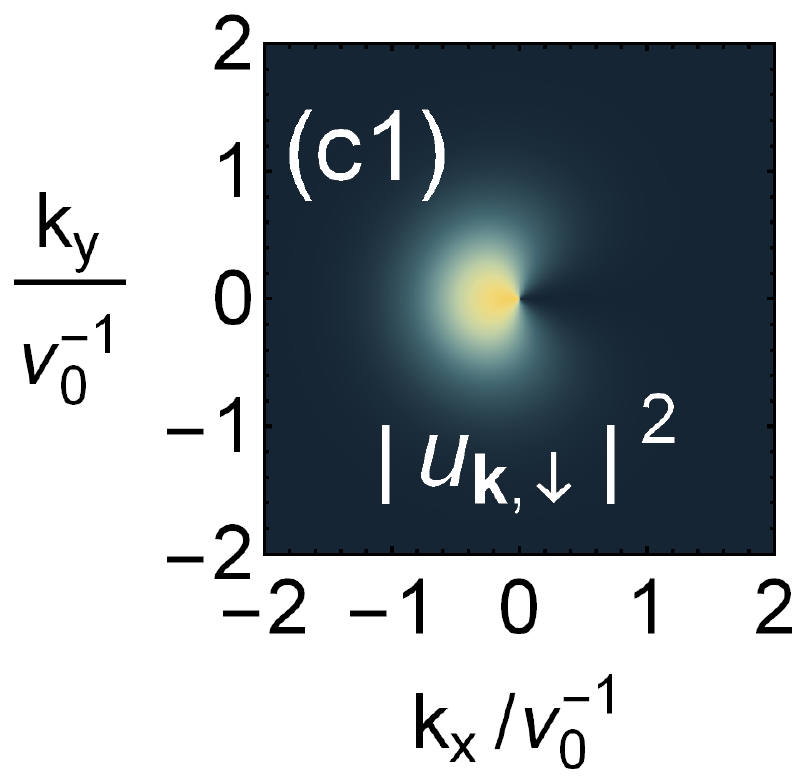}};   
  \node[] at (1.9-2.5, -4.0-0.2-2*\rowsep) { \includegraphics[width=0.106\textwidth,angle=0]{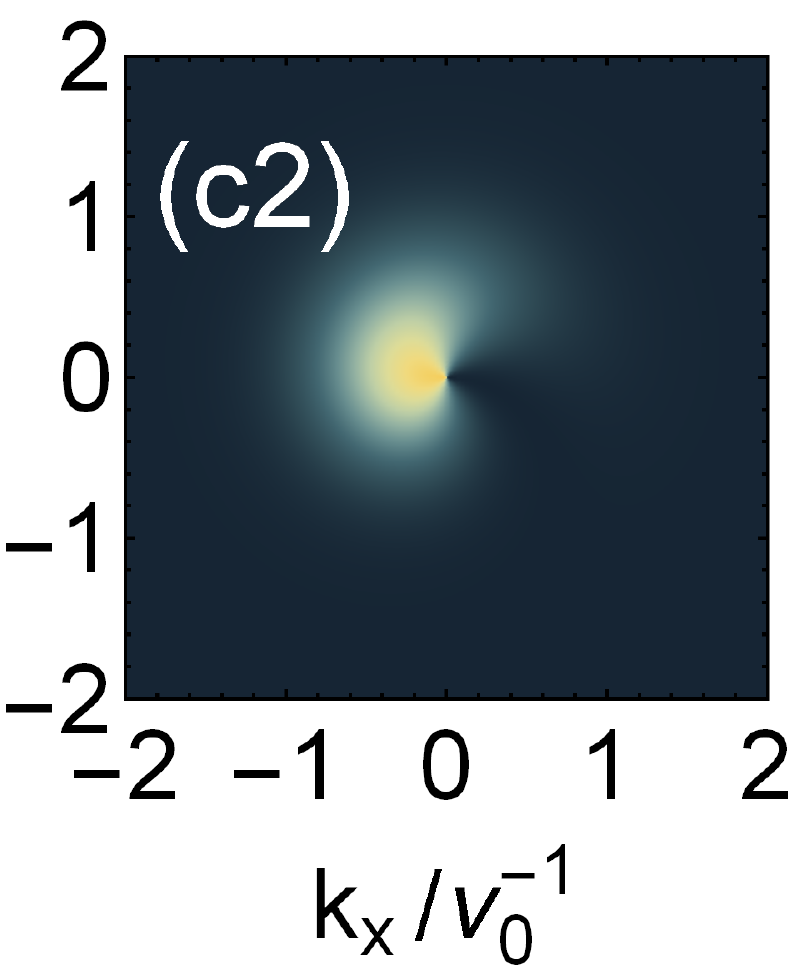}};   
   \node[] at (3.8-2.5, -4.0-0.2-2*\rowsep) { \includegraphics[width=0.106\textwidth,angle=0]{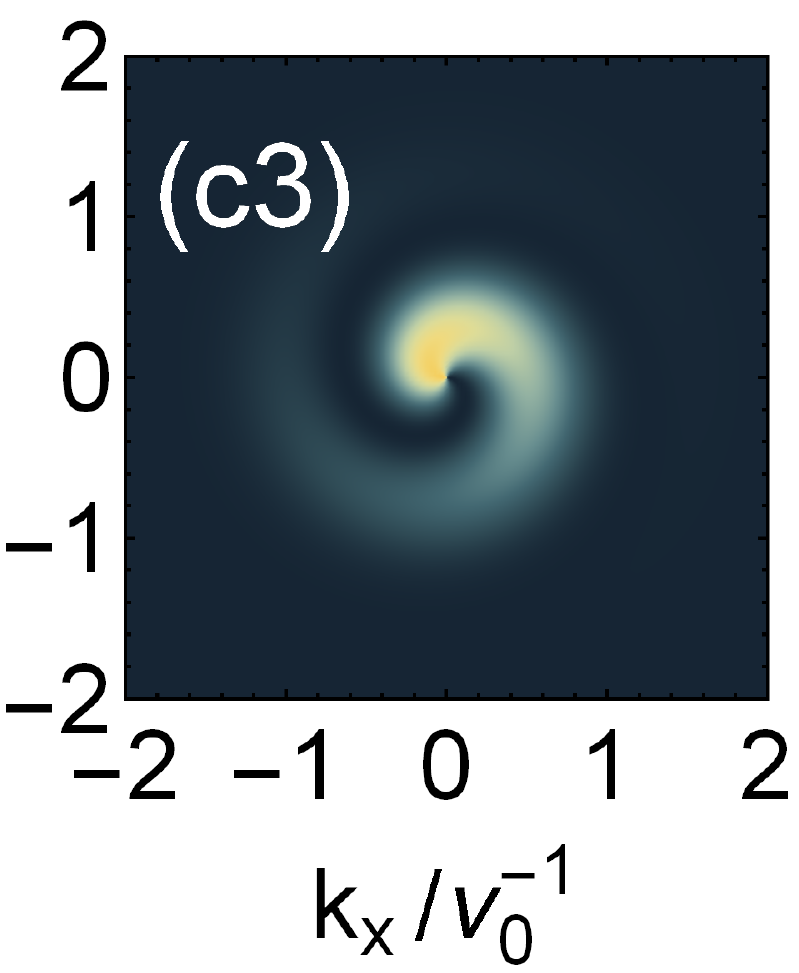}};   
    \node[] at (5.7-2.5, -4.0-0.2-2*\rowsep) { \includegraphics[width=0.106\textwidth,angle=0]{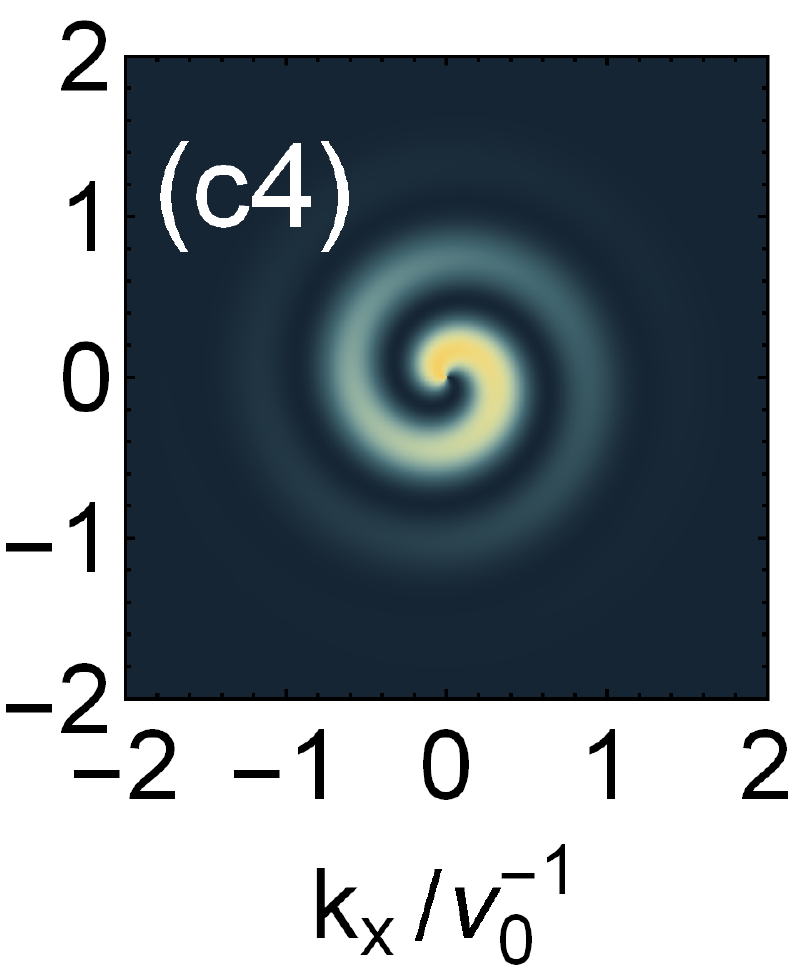}};     
    
     \node[] at (0-2.8, -4.0-0.2-3*\rowsep) { \includegraphics[width=0.13\textwidth,angle=0]{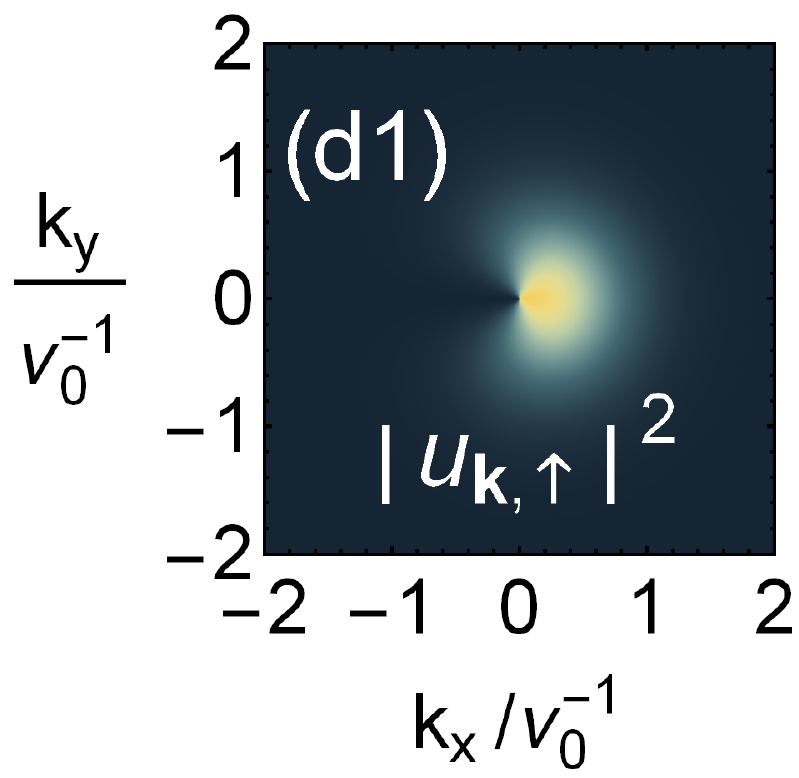}};   
  \node[] at (1.9-2.5, -4.0-0.2-3*\rowsep) { \includegraphics[width=0.106\textwidth,angle=0]{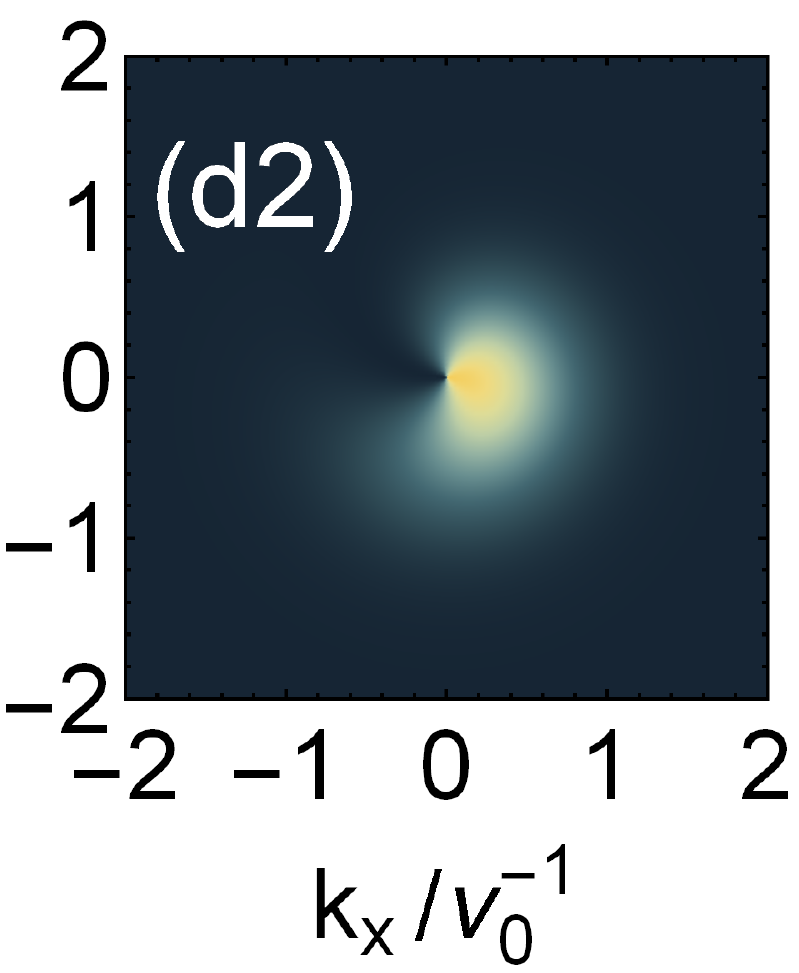}};   
   \node[] at (3.8-2.5, -4.0-0.2-3*\rowsep) { \includegraphics[width=0.106\textwidth,angle=0]{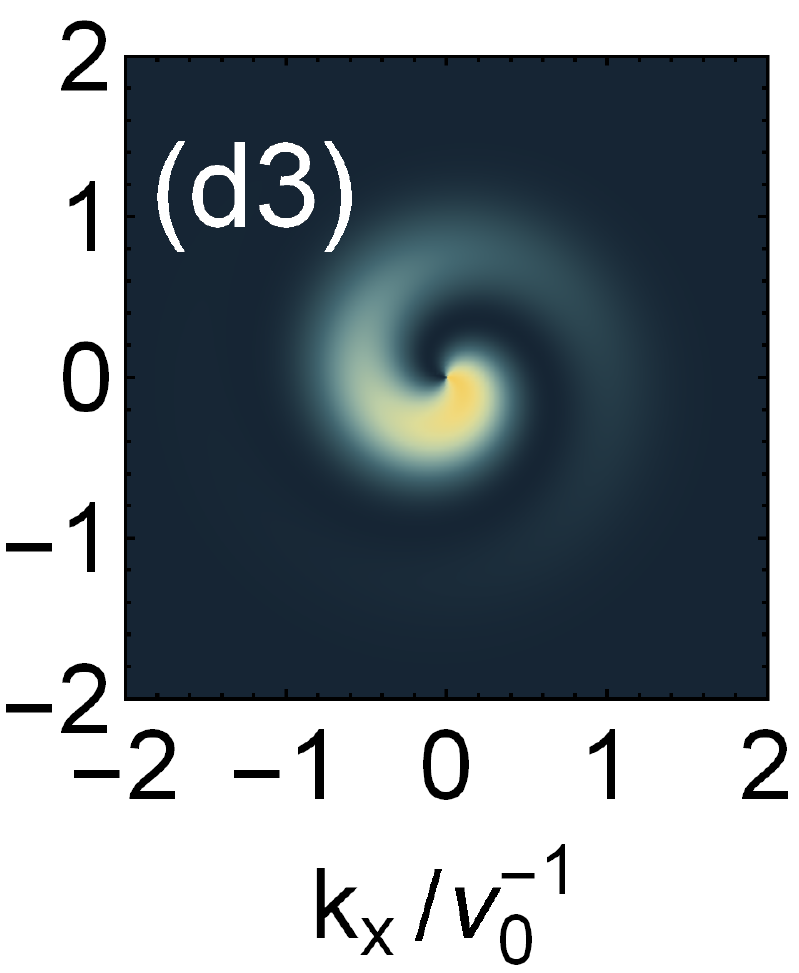}};   
    \node[] at (5.7-2.5, -4.0-0.2-3*\rowsep) { \includegraphics[width=0.106\textwidth,angle=0]{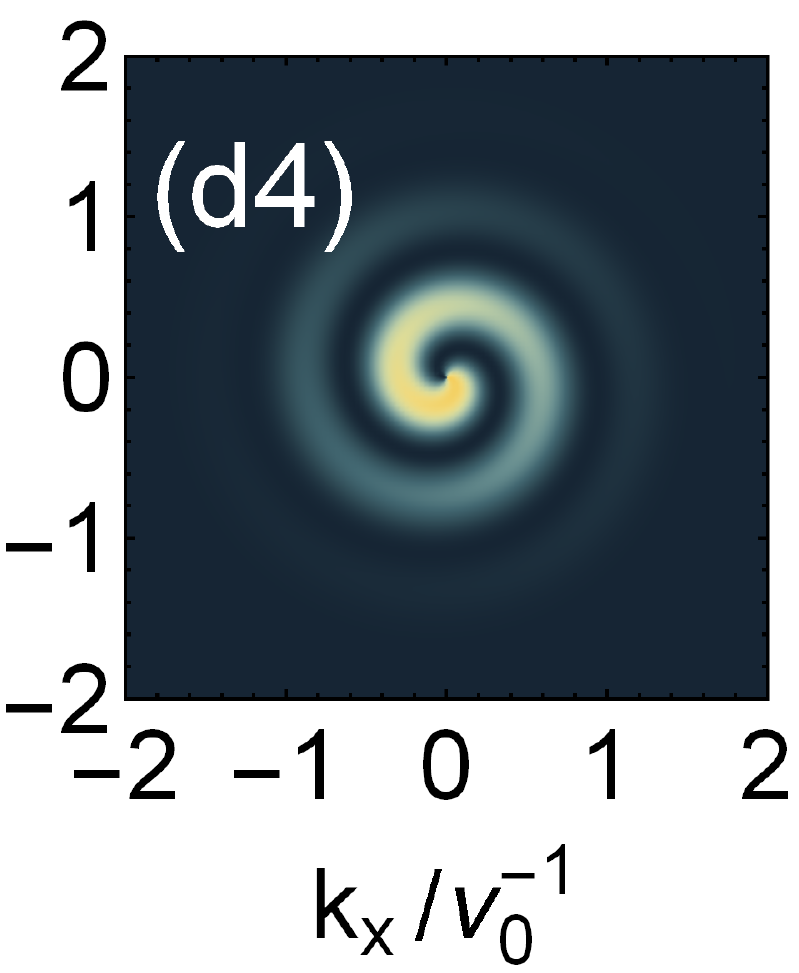}};   
    
\end{tikzpicture}
      \caption{\label{fig:3} Time series of the initial condition shown in Fig.~\ref{fig:2}, using the self-consistent $\Delta$ and $\mu$ shown in Fig.~\ref{fig:selfcons}. (a, b) The phase of $u_{\k,\ua}$, $u_{\k,\da}$ respectively. (c, d) The density $\left|u_{\k,\da} \right|^2$, $\left|u_{\k,\ua} \right|^2$ respectively. The colouring is as in Fig.~\ref{fig:2}a. (1-4) Evaluated at time $t |g|/\hbar = 0, 0.8, 5.0, 10.0$ respectively.}
\end{figure}

Substitution of Eq.~\eqref{eqn:BdGSolnuv-momspace} into Eq.~\eqref{eqn:GapEqHelicityBasis} produces 36 terms, but only 8 of them are non-zero by symmetry. At $t = 0$, due to the initial condition, the gap is given by Eq.~\eqref{eqn:Delta0}, independent of $\lambda$ and $\bar{\mu}$. At later times $t > 0$, we must find the triplet $\{\lambda,\Delta,\bar{\mu}\}(t)$ such that the gap equation, which is complex-valued, is satisfied (that is, the coupled imaginary and real parts must be solved self-consistently). We illustrate in Fig.~\ref{fig:selfcons} one possible set of self-consistent values for the parameters $\{\Delta, \bar{\mu}\}$ as a function of time, with the constraint that $\lambda \nu_0 / |g| = 0.5$ is kept constant. The spin-orbit coupling $\lambda$ need not be a constant, and giving it time dynamics will influence the gap and chemical potential as well (we will not pursue this further here). While it is convenient that a finite order parameter can be found self-consistently for the solution~\eqref{eqn:BdGSolnuv-momspace} in the presence of spin-orbit coupling, the gap equation indicates that $\absD$ decays to zero in the limit $t \to \infty$; therefore, we call this solution a `transient Fermi superfluid'.

Finally, using the numerically obtained self-consistent parameters (Fig.~\ref{fig:selfcons}), we illustrate in Fig.~\ref{fig:3} the time-evolution of the Bogoliubov modes in the basis of the Hamiltonian~\eqref{eqn:HswaveFermiSF-p}. However, simply ignoring the time dynamics, $\Delta(t) = \Delta_0$ and $\bar{\mu}(t) = \bar{\mu}(0)$, does not noticeably change the result. The momentum distribution of the Bogoliubov quasi-particles becomes a spiral whose chirality is determined by the spin projection.

\subsection{Fourier transform of the gap at $t = 0$}

\begin{figure}[t]
\begin{tikzpicture}
   
\pgfmathsetmacro{\rowsep}{2.3}
   
 \node[] at (0,0) { \includegraphics[width=0.24\textwidth,angle=0]{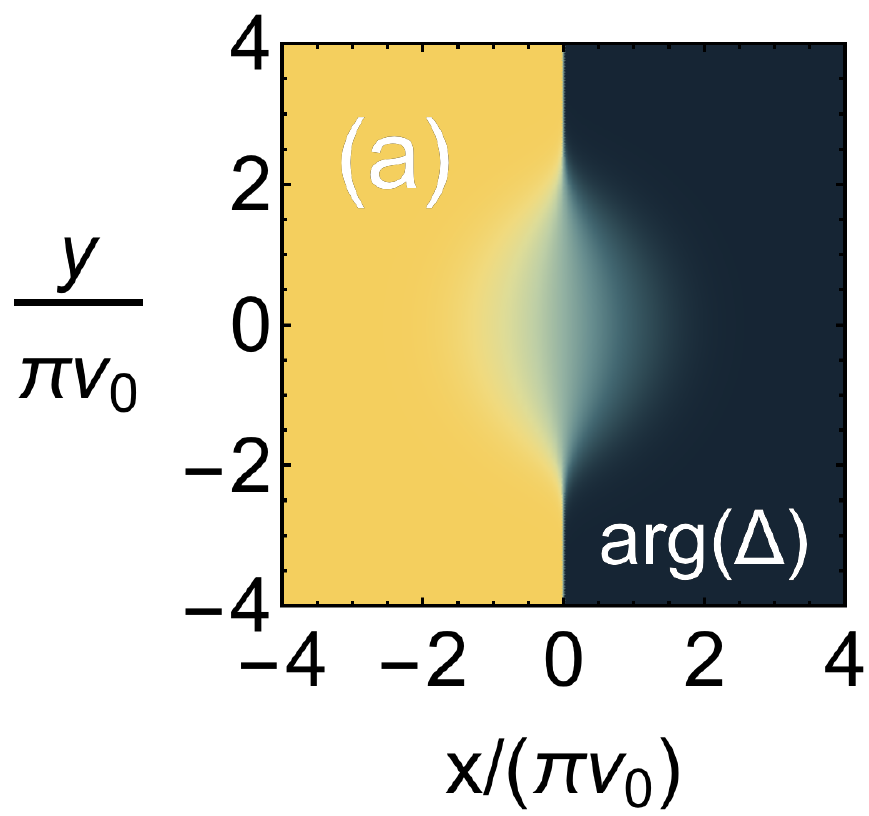}};   
 \node[] at (0.6, 2.2) { \includegraphics[width=0.15\textwidth,angle=0]{2egl}};    
  \node[] at (4.5, 0.0) { \includegraphics[width=0.24\textwidth,angle=0]{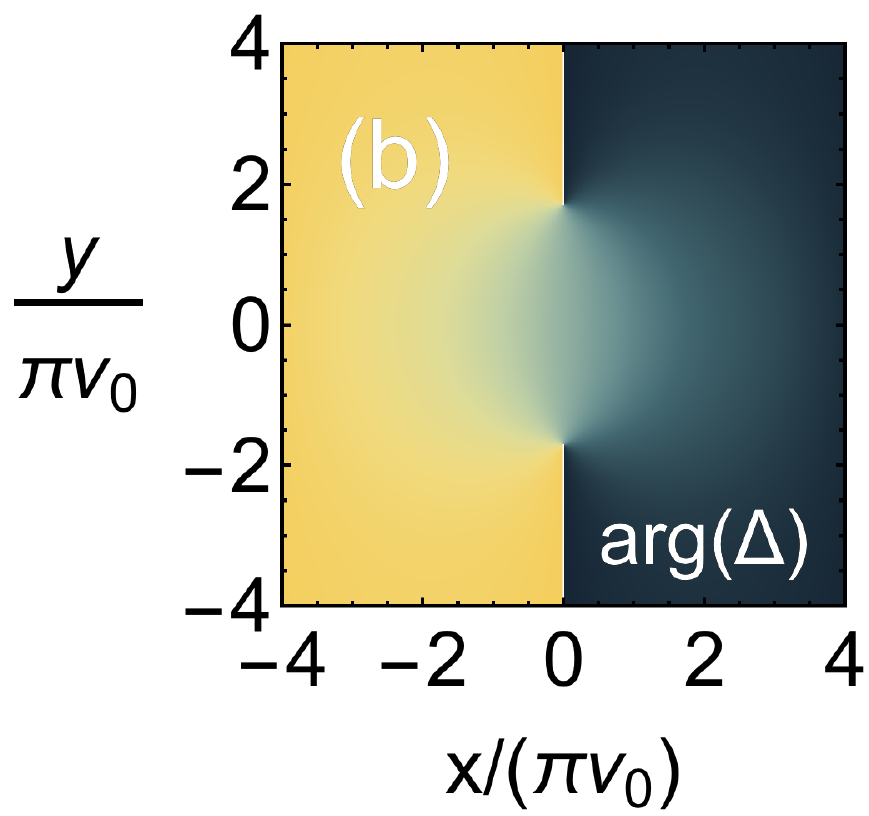}};   
 \node[] at (4.5+0.6, 2.2) { \includegraphics[width=0.15\textwidth,angle=0]{2egl}};  

\end{tikzpicture}
      \caption{\label{fig:5} (a) Analytically and (b) numerically evaluated Fourier transform for obtaining the real-space representation of the superfluid order parameter at $t = 0$.}
\end{figure}

 In momentum space, we have by direct substitution of the initial state into Eq.~\eqref{eqn:GapEqHelicityBasis}
\begin{equation}
\Delta(\k,0) = -\frac{g \nu^2}{\nu_0^2 \pi^3}\rme^{-2 k^2 \nu^2}\left( 1 + \frac{k_\x}{k}\right) = -\frac{g \nu^2}{\nu_0^2 \pi^3}\rme^{-2 k^2 \nu^2}\left[ 1 + \cos{\left(\varphi_\k \right)} \right].
\end{equation}
We will evaluate the gap in real space at $t = 0$ using Eq.~\eqref{eqn:FPINV}:
\begin{equation}
\label{eqn:gapt0realanalytic}
\begin{split}
\Delta(\r,0) &= \nu_0^2 \int \d^2 \k\, \Delta(\k,0) \rme^{-\rmi \k \cdot \r} \\
&=  -\frac{g \nu^2}{  \pi^3} \int_0^\infty k\, \d k \rme^{-2 k^2 \nu^2}  \int_0^{2\pi} \d \varphi_\k\, \left[ 1 + \cos{\left(\varphi_\k \right)} \right] \rme^{-\rmi x k \cos{\left(\varphi_\k\right)} - \rmi y k \sin{\left(\varphi_\k\right)}}\\
&=  -\frac{2g \nu^2}{  \pi^2} \int_0^\infty k\, \d k \rme^{-2 k^2 \nu^2}  \left[ J_0\left( r k \right) - \frac{\rmi x }{ r} J_1\left( r k \right)\right]\\
&=  -\frac{g }{ 2 \pi^2}   \left\lbrace \rme^{-\frac{r^2}{8 \nu^2}}   - \sqrt{\frac{\pi}{2}} \frac{ \rmi x}{4 \nu}  \rme^{-\frac{r^2}{16 \nu^2}} \left[I_0\left(\frac{r^2}{16 \nu^2} \right) - I_1\left(\frac{r^2}{16 \nu^2} \right)    \right]
 \right\rbrace,
\end{split}
\end{equation}
where we used the identity
\begin{equation}
\begin{split}
\int_0^{2 \pi}\cos{\left(m \theta \right)} \rme^{p \cos{\left(\theta\right)} + q \sin{\left(\theta\right)}} \d \theta 
&=  \pi \left(p^2 + q^2 \right)^{-\frac{m}{2}}
\left\lbrace 
\left(p + \rmi q \right)^{m}  + \left(p - \rmi q \right)^{m}
\right\rbrace I_m\left(\sqrt{p^2 + q^2} \right) .
\end{split}
\end{equation}
We compare the analytic and numerical evaluations of the Fourier transform in Fig.~\ref{fig:5}. The finite-time gap in Fig.~\ref{fig:4} is evaluated numerically.

\subsection{Time evolution of the gap}
\begin{figure}[t]
\begin{tikzpicture}
\pgfmathsetmacro{\pwidth}{0.75}
\pgfmathsetmacro{\rowinc}{-4.3}

 \node[] at (0, 0) { \includegraphics[width=\pwidth\textwidth,angle=0]{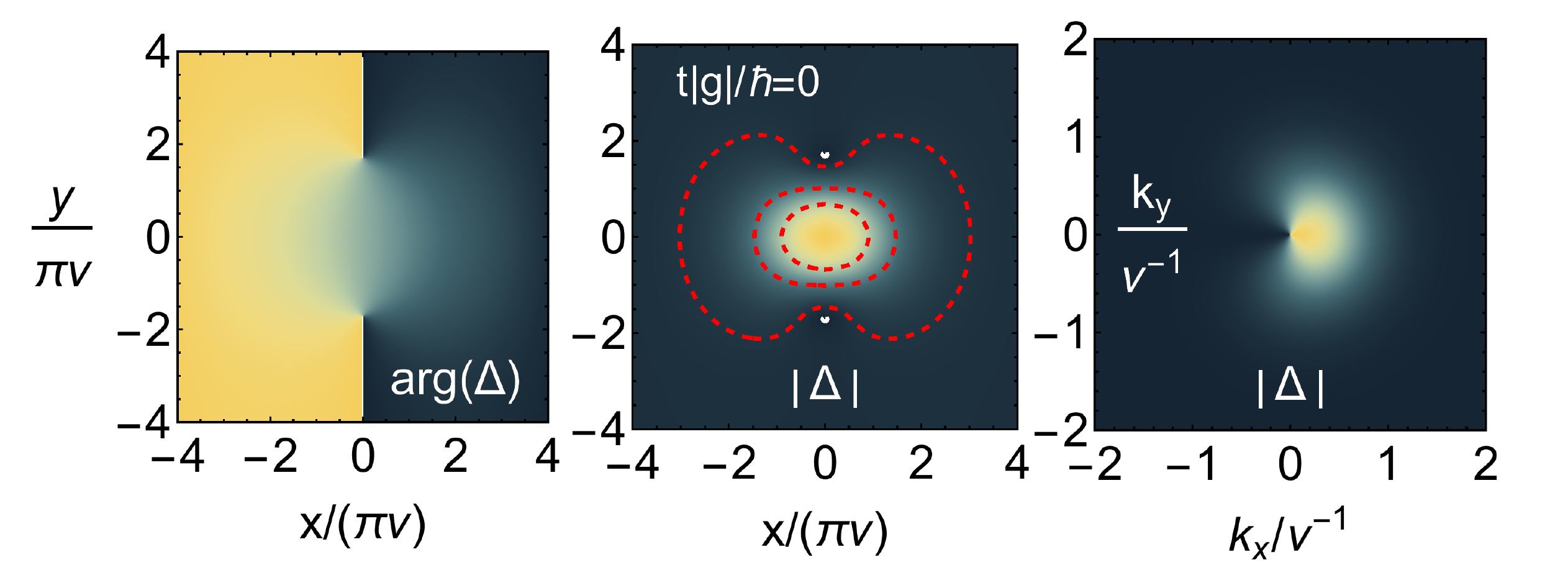}};     
 \node[] at (0, 1*\rowinc) { \includegraphics[width=\pwidth\textwidth,angle=0]{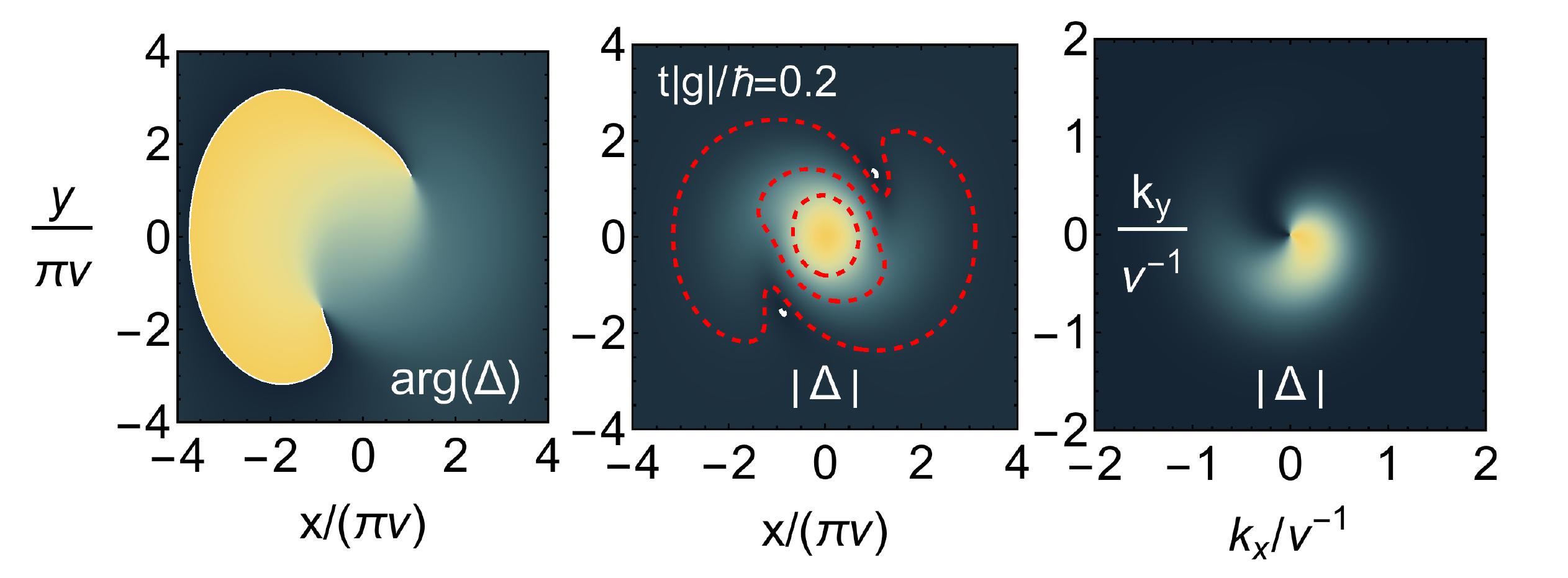}};    
 \node[] at (0, 2*\rowinc) { \includegraphics[width=\pwidth\textwidth,angle=0]{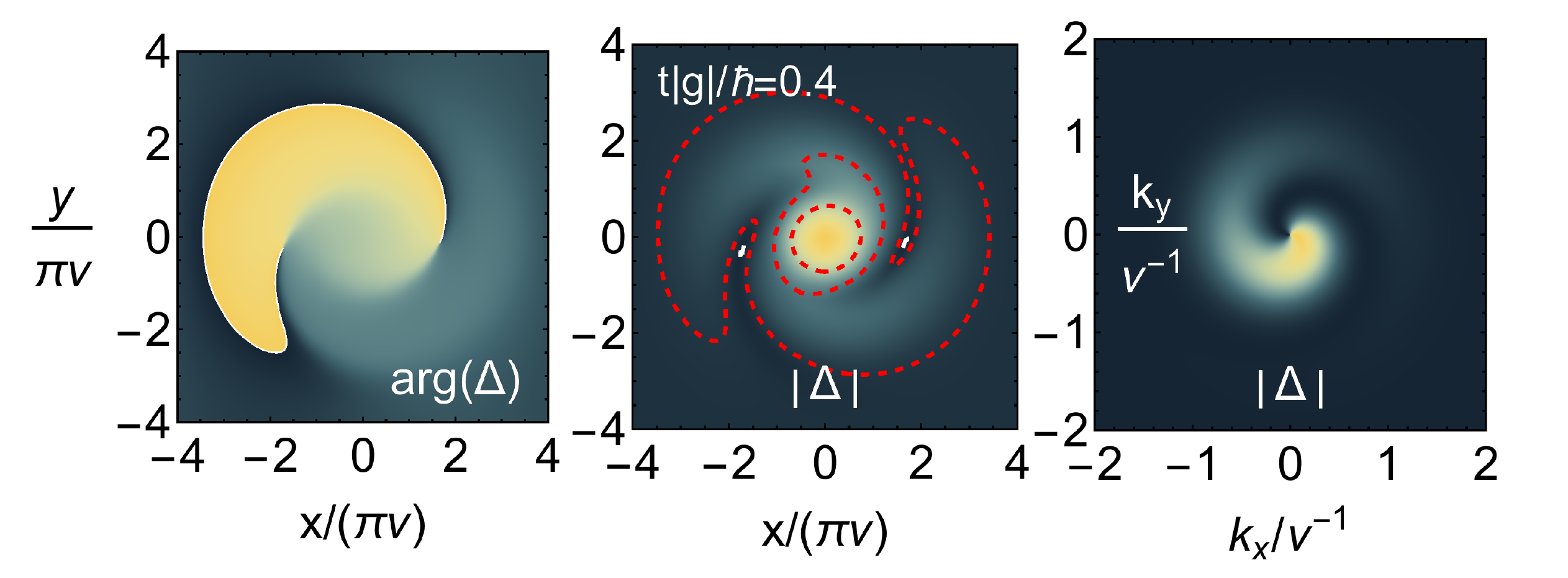}};   
 \node[] at (0, 3*\rowinc) { \includegraphics[width=\pwidth\textwidth,angle=0]{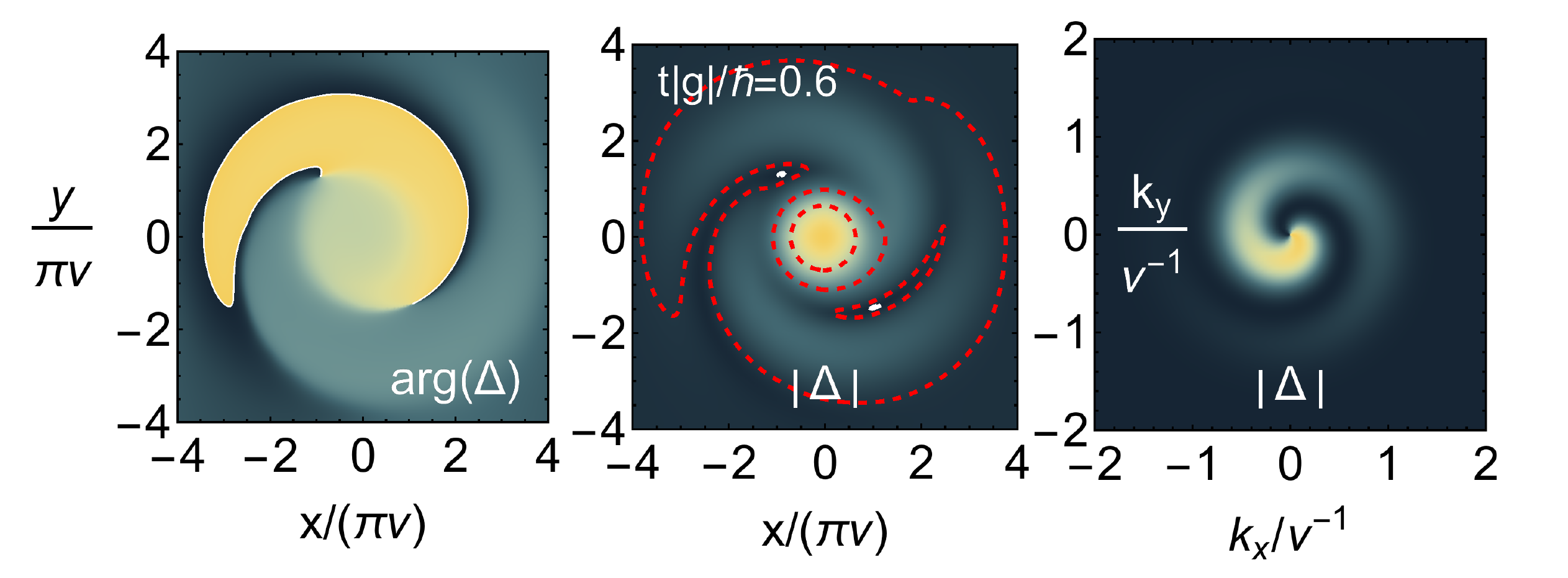}};    
 \node[] at (0, 4*\rowinc) { \includegraphics[width=\pwidth\textwidth,angle=0]{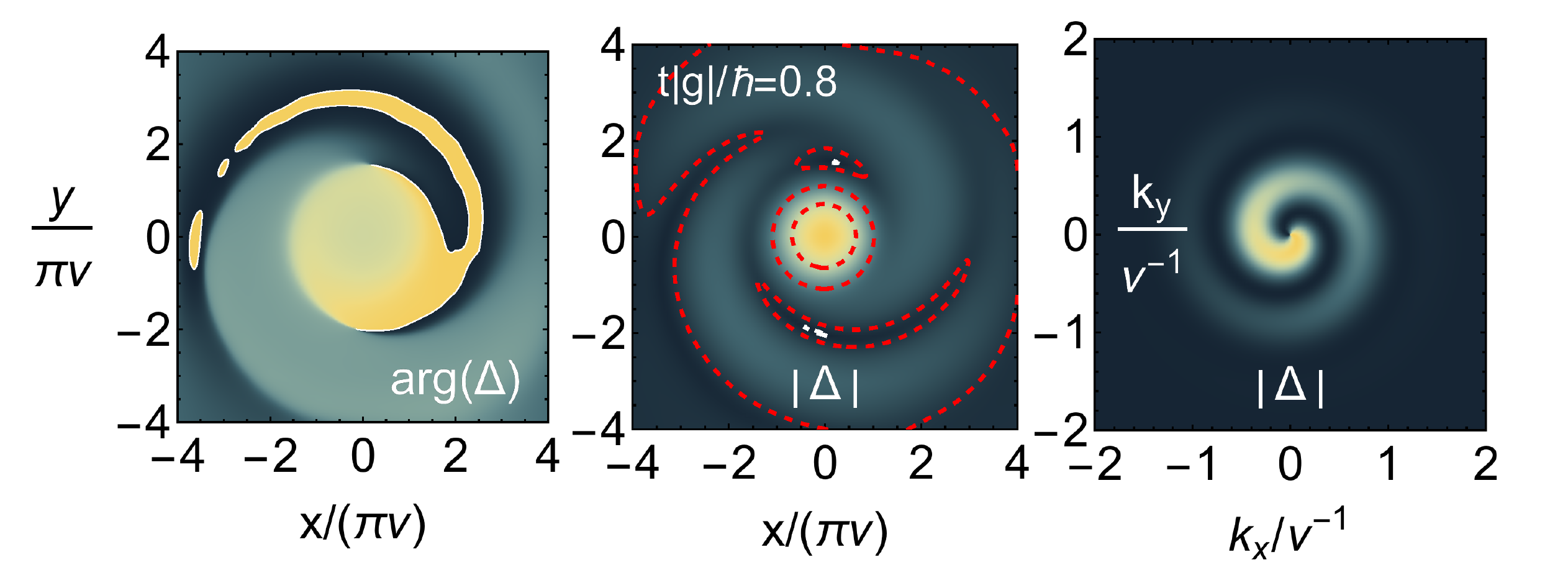}};   
 
    \node[] at (-3.8, 2.5) { \includegraphics[width=0.17\textwidth,angle=0]{2egl}}; 
   \node[] at (0+0.5, 2.6) { \includegraphics[width=0.2\textwidth,angle=0]{densleg}};  
   \node[] at (4.3+0.2, 2.6) { \includegraphics[width=0.2\textwidth,angle=0]{momdensleg}};

\end{tikzpicture}
      \caption{\label{fig:6} The superfluid order parameter $\Delta$ as a function of time for fixed $\lambda \nu /|g|= 5.0$ and $\mu / |g| = 1.0$ (here $\nu_0 = \nu$). The real-space vortex dipole precesses as a result of the Rashba SO coupling $\lambda$. Changing $\lambda \to - \lambda$ equals to reflection of all three colums about the $x$-axis, and the precession direction is reversed.}
\end{figure}

We show the time-evolution of the gap for fixed $\lambda \nu /|g|= 5.0$ and $\mu / |g| = 1.0$ in Fig.~\ref{fig:6}.

\section{\label{sec:app_HelTF}Transformation into the helicity basis}
We denote the different basis representations of the Hamiltonian $\mathcal{H}$ by
\begin{equation}
\begin{split}
\mathcal{H} &= \sum_\kk \, \begin{pmatrix} \textbf{c}_\kk^\dagger & \textbf{c}_{-\kk} \end{pmatrix} H(\kk) \begin{pmatrix} \textbf{c}_\kk & \textbf{c}_{-\kk}^\dagger \end{pmatrix}^\mathrm{T} \\
&= \sum_\kk \, \begin{pmatrix} \textbf{h}_\kk^\dagger & \textbf{h}_{-\kk} \end{pmatrix} H^\mathrm{H}(\kk) \begin{pmatrix} \textbf{h}_\kk & \textbf{h}_{-\kk}^\dagger \end{pmatrix}^\mathrm{T},
\end{split}
\end{equation}
where $\textbf{h}_\k = \begin{pmatrix}
h_{\k,+} & h_{\k,-}
\end{pmatrix}$ and $\textbf{c}_\k = \begin{pmatrix}
c_{\k,\ua} & c_{\k,\da}
\end{pmatrix}$.
We define the `helicity basis' by the relations
\begin{subequations}
\label{eqn:HelTFdefs}
\begin{align}
\varphi_\k &= \mathrm{arg}\left(k_\x + \rmi k_\y \right),\\
\varphi_{-\k} &= \varphi_{\k}+\pi,\\
k \rme^{\pm \rmi \varphi_\k} &= k_\x \pm \rmi k_\y,\\
\label{eqn:heltfdefd}
\mathcal{T}_\k &= \frac{1}{\sqrt{2}}
\begin{pmatrix}
1 & \rme^{\rmi \varphi_\k}\\
\rme^{-\rmi \varphi_\k} & -1
\end{pmatrix} = \mathcal{T}_\k^{-1} = \mathcal{T}_\k^{\dagger} ,\\
\label{eqn:heltfdefe}
\mathcal{T}_{-\k}^\mathrm{T} &= \frac{1}{\sqrt{2}}
\begin{pmatrix}
1 & -\rme^{-\rmi \varphi_\k}\\
-\rme^{\rmi \varphi_\k} & -1
\end{pmatrix} = \left( \mathcal{T}_{-\k}^\mathrm{T} \right)^{-1} = \left(\mathcal{T}_{-\k}^\mathrm{T}\right)^\dagger,\\
\begin{pmatrix}
h_{\k,+}\\
h_{\k,-}
\end{pmatrix}
&= 
\mathcal{T}_\k
\begin{pmatrix}
c_{\k,\ua}\\
c_{\k,\da}
\end{pmatrix},\\
\begin{pmatrix}
h_{\k,+}^\dagger &
h_{\k,-}^\dagger
\end{pmatrix}
&= 
\begin{pmatrix}
c_{\k,\ua}^\dagger &
c_{\k,\da}^\dagger
\end{pmatrix}
\mathcal{T}_\k^\dagger,\\
\begin{pmatrix}
h_{-\k,+}^\dagger\\
h_{-\k,-}^\dagger
\end{pmatrix}
&= 
\mathcal{T}_{-\k}^\mathrm{T}
\begin{pmatrix}
c_{-\k,\ua}^\dagger \\
c_{-\k,\da}^\dagger
\end{pmatrix}, \\
\begin{pmatrix}
h_{-\k,+} &
h_{-\k,-}
\end{pmatrix}
&= 
\begin{pmatrix}
c_{-\k,\ua} &
c_{-\k,\da}
\end{pmatrix}
\left(\mathcal{T}_{-\k}^\mathrm{T}\right)^\dagger
\end{align}
\end{subequations}
With the definitions~\eqref{eqn:HelTFdefs}, the Hamiltonian defined in Eq.~\eqref{eqn:HswaveFermiSF-p} transforms as
\begin{equation}
\label{eqn:HswaveFermiSF-papp}
\begin{split}
H^\mathrm{H}(\kk) &= \begin{pmatrix}
 \mathcal{T}_\k^\dagger \left( H_0 - h \sigma_\mathrm{z} + \lambda \textbf{g}_\kk \cdot \boldsymbol{\sigma} \right) \mathcal{T}_\k&  \mathcal{T}_\k^\dagger \left(-\rmi \, \Delta \sigma_\mathrm{y} \right) \mathcal{T}_{-\k}^\mathrm{T} \\\left(\mathcal{T}_{-\k}^\mathrm{T}\right)^\dagger
\left(\rmi \, \Delta^* \sigma_\mathrm{y}\right) \mathcal{T}_\k & \left(\mathcal{T}_{-\k}^\mathrm{T}\right)^\dagger\left( -H_0 + h \sigma_\mathrm{z} + \lambda \textbf{g}_\kk \cdot \boldsymbol{\sigma}^* \right) \mathcal{T}_{-\k}^\mathrm{T}  \\
\end{pmatrix}\\
&= \begin{pmatrix}
H_0 + k \lambda & - \rme^{\rmi \varphi_\k} h & \rme^{\rmi \varphi_\k} \Delta & 0\\
- \rme^{-\rmi \varphi_\k} h & H_0 - k \lambda & 0 & \rme^{-\rmi \varphi_\k} \Delta \\
\rme^{-\rmi \varphi_\k} \Delta^* & 0& -H_0 - k \lambda & - \rme^{-\rmi \varphi_\k} h\\
0 & \rme^{\rmi \varphi_\k} \Delta^* & - \rme^{\rmi \varphi_\k} h & -H_0 + k \lambda\\
\end{pmatrix},
\end{split}
\end{equation}
which is Eq.~\eqref{eqn:HswaveFermiSF-p-HelTF} in the main text.

\section{\label{sec:app_BdGeval}Solution of the time-dependent BdG equations}
Any complex $2\times 2$ matrix can be parametrised by the vector $\B = \begin{pmatrix}
B_1 & B_2 & B_3
\end{pmatrix}^\mathrm{T}$ as $B_0 \sigma_0 + \begin{pmatrix}
\sigma_\x & \sigma_\y & \sigma_\z
\end{pmatrix}^\mathrm{T} \cdot \B$. We now drop $B_0 = 0$. For $H^\mathrm{H}_\pm$, taking $\Delta$ to be real, we have $\B^\mathrm{H} = \begin{pmatrix}
\frac{\Delta}{k}k_\x & \mp \frac{\Delta}{k}k_\y & \epsilon_\k - \bar{\mu} \pm \lambda k
\end{pmatrix}^\mathrm{T}$. The square $\left(H^\mathrm{H}_\pm\right)^2$ is diagonal. This results in the general observation that 
\begin{equation}
\exp{\left[-\frac{\rmi}{\hbar}H^\mathrm{H}_\pm t \right]} = \cos{\left(\frac{\absB t}{\hbar}  \right)}\sigma_0 - \rmi \sin{\left(\frac{\absB t}{\hbar}  \right)} \frac{H^\mathrm{H}_\pm }{\absB}.
\end{equation}
Here $\absB = \sqrt{\left(\epsilon_\k - \bar{\mu} \pm \lambda k \right)^2  + \absD^2 }$ corresponds to the magnitude of the eigenvalues of $H^\mathrm{H}_\pm$, which coincide with the eigenvalues of $H(\k)$. The time evolution is given by $\phi^\pm_\k(t) = \exp{\left[-\frac{\rmi}{\hbar}H^\mathrm{H}_\pm t \right]} \phi^\pm_\k(0)$, where $\phi^\pm_\k(0) = \begin{pmatrix} \phi^\pm_0(\k) & \chi^\pm_0(\k) \end{pmatrix}^\mathrm{T}$, which yields
\begin{subequations}
\begin{align}
u^\pm_{\k,\eta_0}(t) &= \left[\cos{\left(\frac{\absB t}{\hbar}  \right)}  - \rmi  \sin{\left(\frac{\absB t}{\hbar}  \right)} \frac{\left(\epsilon_\k - \bar{\mu} \pm \lambda k \right) }{\absB}\right]\phi_0^\pm(\k)- \rmi  \sin{\left(\frac{\absB t}{\hbar}  \right)} \frac{\frac{\Delta}{k} \left(k_\x \pm \rmi k_\y \right)}{\absB} \chi^\pm_0(\k),\\
v^\pm_{\k,\eta_0}(t) &= - \rmi  \sin{\left(\frac{\absB t}{\hbar}  \right)} \frac{\frac{\Delta^*}{k} \left(k_\x \mp \rmi k_\y \right) }{\absB} \phi_0^\pm(\k) + 
\left[\cos{\left(\frac{\absB t}{\hbar}  \right)}   - \rmi  \sin{\left(\frac{\absB t}{\hbar}  \right)} \frac{-\left(\epsilon_\k - \bar{\mu} \pm \lambda k \right) }{\absB}\right] \chi_0^\pm(\k).  
\end{align}
\end{subequations}
This is an exact result.

We take $\phi_0^\pm(\k) = \frac{\nu}{\nu_0 \sqrt{2\pi}\pi}\,\rme^{-\nu^2 k^2}$ as the Gaussian initial condition. We now focus on the low-energy spectrum also taking $\lambda = 0$ and $\chi^\pm_0(\k) = 0$ for simplicity.  When $\k$  is small, we can write $\absB \simeq \sqrt{\bar{\mu}^2  + \absD^2 }$, and importantly $k$-independent. We can then evaluate $\mathcal{F}_\r^{-1}$, the dimensionless Fourier transformation evaluated at position $\r$. Our definitions are as follows:
\begin{subequations}
\begin{align}
\label{eqn:FPFWD}
\mathcal{F}_\k\left[f(\r)\right] &= \frac{1}{(2\pi)^2}  \frac{1}{\nu_0^2} \int \d^2 \r\, f(\r) \rme^{\rmi \k \cdot \r},\\
\label{eqn:FPINV}
\mathcal{F}_\r^{-1}\left[g(\k)\right]&= \nu_0^2 \int \d^2 \k\, g(\k) \rme^{-\rmi \k \cdot \r}.
\end{align}
\end{subequations}
By inspection
\begin{equation}
\begin{split}
\label{eqn:urnoFPTF}
u_{\r,\eta_0}^\pm(t) &= \cos{\left(\frac{\absB t}{\hbar}  \right)}\phi^\pm_0(\r)  + \rmi \frac{\bar{\mu}}{\absB}  \sin{\left(\frac{\absB t}{\hbar}  \right)}\phi^\pm_0(\r).
\end{split}
\end{equation}
For the other component, we then have
\begin{equation}
\begin{split}
\label{eqn:vrnoFPTF}
v_{\r,\eta_0}^\pm(t) &= \underbrace{- \rmi   \frac{\nu \nu_0 \sin{\left(\absB t/\hbar \right)} \Delta^* }{\pi\sqrt{2\pi}\absB}}_{\equiv \Gamma} \int \d^2 k  \frac{k_\x \mp \rmi k_\y}{k}  \exp{\left[-\nu^2 k^2 
- \rmi \k \cdot \r  \right]}\\
 &= \Gamma \int_0^\infty \d k\, k \exp{\left[- \nu^2  k^2 
 \right]}
 \int_0^{2\pi} \d \varphi_\k  \rme^{\mp \rmi \varphi_\k} \rme^{-\rmi x k \cos{\left(\varphi_\k \right)}  - \rmi y k \sin{\left(\varphi_\k \right)} }.
\end{split}
\end{equation}
To evaluate the integral over $\varphi_\k$, we consider the following identity:
\begin{equation}
\begin{split}
\label{eqn:idGR3.937}
\int_0^{2 \pi}\rme^{\pm \rmi m \theta } \rme^{p \cos{\left(\theta\right)} + q \sin{\left(\theta\right)}} \d \theta
 &=  2\pi \left(p^2 + q^2 \right)^{-\frac{m}{2}}
 \left(p^2 - q^2 \pm \rmi 2 pq \right)^{\frac{m}{2}}I_m\left(\sqrt{p^2 + q^2} \right),
\end{split}
\end{equation}
where $m = 0,1,2,\ldots$. We have in general $p = (\pm)\rmi x k$, $q = (\pm)\rmi y k$, and $m = 1$. Therefore $p^2 + q^2 = -r^2 k^2$, $p^2 - q^2 \pm \rmi 2 pq = k^2 \left(- x^2 + y^2  \mp \rmi 2 xy \right) = k^2  \left(y \mp \rmi x\right)^2$, and
\begin{equation}
 \int_0^{2\pi} \d \varphi_\k  \rme^{\mp \rmi \varphi_\k} \rme^{-\rmi x k \cos{\left(\varphi_\k \right)}  - \rmi y k \sin{\left(\varphi_\k \right)} } =  2\pi \frac{\left( \left(y \pm \rmi x\right)^2 \right)^{\frac{1}{2}}}{\left(-r^2  \right)^{\frac{1}{2}}} 
 I_1\left(\sqrt{-r^2 k^2} \right)  
 =  2\pi \frac{y \pm \rmi x}{r} 
 J_1\left( rk \right) 
\end{equation}
if $r,k\geq 0$.
Then
\begin{equation}
\begin{split}
v^\pm_{\r,\eta_0}(t) 
 &= \Gamma   2\pi \frac{y \pm \rmi x}{r} \int_0^\infty \d k\, k \exp{\left[- \nu^2  k^2 
 \right]} 
 J_1\left( rk \right). 
\end{split}
\end{equation}
An exact solution is possible for the first-order Hankel transform using the table integral
\begin{equation}
\label{eqn:idGR6.631}
\int_0^\infty x^\mu \rme^{-\alpha x^2} J_\nu (\beta x) \d x = \frac{ \Gamma\left(\frac{\nu}{2} + \frac{\mu}{2}+ \frac{1}{2} \right)}{\beta \alpha^{\frac{\mu}{2}}\Gamma\left( \nu + 1 \right)}  \exp{\left(-\frac{\beta^2}{8\alpha} \right)} M_{\frac{\mu}{2},\frac{\nu}{2}}\left(\frac{\beta^2}{4\alpha} \right), \qquad \mathrm{Re}(\alpha) > 0, \mathrm{Re}(\mu + \nu ) > -1,
\end{equation}
where $M$ is the Whittaker $M$-function. Then
\begin{equation}
\begin{split}
 \int_0^\infty \d k\, k \exp{\left(-\nu^2 k^2 \right)}
  J_1\left( rk \right) 
  &=  \frac{\sqrt{\pi }}{8 \nu^3} r \rme^{-\frac{r^2}{8 \nu^2}} \left[I_0\left(\frac{r^2}{8 \nu^2}\right)-I_1\left(\frac{r^2}{8 \nu^2}\right)\right],
\end{split}
\end{equation}
and
\begin{equation}
\begin{split}
v^\pm_{\r,\eta_0}(t) 
 &=\left(- \rmi y \pm x\right)  \frac{ \nu \nu_0 \sin{\left(\absB t/\hbar \right)} \Delta^* }{4\sqrt{2}\absB  \nu^3}     \rme^{-\frac{r^2}{8 \nu^2}} \left[I_0\left(\frac{r^2}{8 \nu^2}\right)-I_1\left(\frac{r^2}{8 \nu^2}\right)\right]. 
\end{split}
\end{equation}
There is no dispersion of the wavepacket ($\nu \in \mathbb{R}$) because we have ignored $\epsilon_\k$. Including the kinetic term makes evaluating the Fourier transform cumbersome, and is not an essential feature for observation of the helicity vortex itself.

\section{\label{app:dSsol} Solution for the Hamiltonian~\eqref{eqn:matH2x2pm-dasSarma}}
In momentum space, let us denote $K_0 = \hat{T} + \hat{V}$, where 
\begin{subequations}
\begin{align}
\hat{T} &= \left(\epsilon_\kk - \bar{\mu}\right)\sigma_0 + h \sigma_\z,\\ 
\hat{V} &= \begin{pmatrix}
0 & -\lambda \left(k_\y - \rmi k_\x \right) \\
-\lambda \left(k_\y + \rmi k_\x \right)& 0
\end{pmatrix},
\end{align}
\end{subequations}
This implies $\left[\hat{T}, \hat{V}\right] \propto  \lambda h$. If $h = 0$, then using the Baker-Campbell-Hausdorff formula we can write $ \rme^{-\frac{\rmi}{\hbar}  K_0 t } =  \rme^{-\frac{\rmi}{\hbar}  \hat{T} t } \rme^{-\frac{\rmi}{\hbar}  \hat{V} t }$, where
\begin{subequations}
\begin{align}
 \rme^{-\frac{\rmi}{\hbar}  \hat{T} t } &= \exp{\left[-\frac{\rmi}{\hbar}\left(\epsilon_\kk - \bar{\mu} \right) t \sigma_0 \right]},\\
 \rme^{-\frac{\rmi}{\hbar} \hat{ V} t } &= \cos{\left(\frac{k \lambda  t}{\hbar}  \right)}\sigma_0 - \rmi \sin{\left(\frac{k \lambda  t}{\hbar}  \right)} \frac{\hat{V}_\pm}{k \lambda },
\end{align}
\end{subequations}
because $\hat{V}^2 = k^2 \lambda^2 \sigma_0$. More generally, finite Zeeman field results in additional terms for the decomposition of $ \rme^{-\frac{\rmi}{\hbar}  K_0 t }$ as a power series in $h$. We now seek a solution to $K_0 \begin{pmatrix}
q_\r(t) \\
w_\r(t)
\end{pmatrix} = \rmi \hbar \partial_t \begin{pmatrix}
q_\r(t) \\
w_\r(t)
\end{pmatrix} $ with the initial condition $q_\r(0) = \phi_0^\pm(\r)$, $w_\r(0) = 0$.

\subsection{\label{sec:app_BdGeval1w}Evaluation of $w_{\r}(t)$}
We have
\begin{equation}
\begin{split}
\label{eqn:wrnoFPTF}
w_\r(t) &= \underbrace{  \frac{\lambda}{\lambda } \frac{\nu \nu_0  \exp{\left(\frac{\rmi}{\hbar}\bar{\mu} t\right)}  }{\pi\sqrt{2\pi}}}_{\equiv \Gamma} \int \d^2 k  \sin{\left(k \lambda  t/\hbar \right)}\frac{k_\x - \rmi k_\y}{k}  \exp{\left[-\left( \nu^2 + \frac{\rmi  \hbar t}{2m} \right) k^2 
- \rmi \k\cdot \r  \right]}\\
 &= \Gamma \int_0^\infty \d k\, k \sin{\left(k \lambda  t/\hbar \right)} \exp{\left[-A_+ k^2 
 \right]}
 \int_0^{2\pi} \d \varphi_\k  \rme^{- \rmi \varphi_\k} \rme^{-\rmi x k \cos{\left(\varphi_\k \right)}  - \rmi y k \sin{\left(\varphi_\k \right)} },
\end{split}
\end{equation}
where $A_+ = \nu^2 + \frac{\rmi  \hbar t}{2m} $. Using the definition~\eqref{eqn:FPINV}, Eq.~\eqref{eqn:wrnoFPTF} is simply the inverse Fourier transform of $\Gamma \left(k_\x - \rmi k_\y\right) \sin{\left(k \lambda  t/\hbar \right)} \rme^{-A_+ k^2}$. To evaluate the integral over $\varphi_\k$, we consider the identity~\eqref{eqn:idGR3.937}. We have 
\begin{equation}
 \int_0^{2\pi} \d \varphi_\k  \rme^{- \rmi \varphi_\k} \rme^{-\rmi x k \cos{\left(\varphi_\k \right)}  - \rmi y k \sin{\left(\varphi_\k \right)} } =  2\pi \frac{\left( \left(y + \rmi x\right)^2 \right)^{\frac{1}{2}}}{\left(-r^2  \right)^{\frac{1}{2}}} 
 I_1\left(\sqrt{-r^2 k^2} \right)  
 =  2\pi \frac{y + \rmi x}{r} 
 J_1\left( rk \right) 
\end{equation}
if $r,k\geq 0$.
Then
\begin{equation}
\begin{split}
w_\r(t) 
 &=  \Gamma 2\pi \frac{y + \rmi x}{r}   
 \int_0^\infty \d k\, k \sin{\left(k \lambda  t/\hbar \right)} \exp{\left[-A_+ k^2  \right]}
  J_1\left( rk \right) .
\end{split}
\end{equation}
An exact solution is possible for the first-order Hankel transform using the table integral~\eqref{eqn:idGR6.631}:
\begin{equation}
\begin{split}
 \int_0^\infty \d k\, k \sin{\left(k \lambda  t/\hbar \right)} \exp{\left[-A_+ k^2  \right]}
  J_1\left( rk \right) &= \sum_{\mu = 0}^\infty \frac{(-1)^\mu}{(2\mu+1)!} \left( \frac{\lambda t}{\hbar}\right)^{2\mu+1} \int_0^\infty k^{2\mu + 2} \rme^{-A_+ k^2} J_1 (r k) \d k \\
  &=\frac{\exp{\left(-\frac{r^2}{8 A_+} \right)}}{r} \sum_{\mu = 0}^\infty  \left( \frac{\lambda t}{\hbar}\right)^{2\mu+1}  \frac{(-1)^\mu \Gamma\left(\mu + \frac{3}{2} \right)}{(2\mu+1)! \, A_+^{\mu + 1}}   M_{\mu +1,\frac{1}{2}}\left(\frac{r^2}{4A_+} \right),
\end{split}
\end{equation}
where $M$ is the Whittaker $M$-function. Then
\begin{equation}
\begin{split}
\label{eqn:functionDr2}
w_\r(t) 
 &=     \frac{2\nu \nu_0  \exp{\left(\frac{\rmi}{\hbar}\bar{\mu} t\right)}  }{\sqrt{2\pi}}  \frac{y +\rmi x }{r^2}   
\exp{\left(-\frac{r^2}{8 A_+} \right)} \sum_{\mu = 0}^\infty  \left( \frac{\lambda t}{\hbar}\right)^{2\mu+1}  \frac{(-1)^\mu \Gamma\left(\mu + \frac{3}{2} \right)}{(2\mu+1)! \, A_+^{\mu + 1}}   M_{\mu +1,\frac{1}{2}}\left(\frac{r^2}{4A_+} \right).
\end{split}
\end{equation}
While the $r$-dependence in the sum is complicated, a similar prefactor, $x - \rmi y$, characterising a singly-quantised vortex appears.

\subsection{\label{sec:app_BdGeval1}Evaluation of $q_{\r}(t)$}
We have
\begin{equation}
\begin{split}
\label{eqn:qrnoFPTF}
q_\r(t) &= \underbrace{ \frac{\nu \nu_0  \exp{\left(\frac{\rmi}{\hbar}\bar{\mu} t\right)}  }{\pi\sqrt{2\pi}}}_{\equiv \Gamma} \int \d^2 k  \cos{\left(k \lambda  t/\hbar \right)}\exp{\left[-\left( \nu^2 + \frac{\rmi  \hbar t}{2m} \right) k^2 
- \rmi \k\cdot \r  \right]}\\
 &= \Gamma \int_0^\infty \d k\,k \cos{\left(k \lambda  t/\hbar \right)} \exp{\left[-A_+ k^2 
 \right]}
 \int_0^{2\pi} \d \varphi_\k  \rme^{-\rmi x k \cos{\left(\varphi_\k \right)}  - \rmi y k \sin{\left(\varphi_\k \right)} },
\end{split}
\end{equation}
where $A_+ = \nu^2 + \frac{\rmi  \hbar t}{2m} $. To evaluate the integral over $\varphi_\k$, we consider the identity~\eqref{eqn:idGR3.937}. We have 
\begin{equation}
 \int_0^{2\pi} \d \varphi_\k  \rme^{-\rmi x k \cos{\left(\varphi_\k \right)}  - \rmi y k \sin{\left(\varphi_\k \right)} } =  2\pi 
 J_0\left( rk \right) 
\end{equation}
if $r,k\geq 0$.
Then
\begin{equation}
\begin{split}
q_\r(t) 
 &=   \Gamma 2\pi 
 \int_0^\infty \d k\, k \cos{\left(k \lambda  t/\hbar \right)} \exp{\left[-A_+ k^2  \right]}
  J_0\left( rk \right) .
\end{split}
\end{equation}
An exact solution is possible for the first-order Hankel transform using the table integral~\eqref{eqn:idGR6.631}:
\begin{equation}
\begin{split}
 \int_0^\infty \d k\,k \cos{\left(k \lambda  t/\hbar \right)} \exp{\left[-A_+ k^2  \right]}
  J_0\left( rk \right) &= \sum_{\mu = 0}^\infty \frac{(-1)^\mu}{(2\mu)!} \int_0^\infty  k^{2\mu + 1}\left( \frac{\lambda t}{\hbar}\right)^{2\mu}  \rme^{-A_+ k^2} J_0 (r k) \d k \\
  &=\frac{\exp{\left(-\frac{r^2}{8 A_+} \right)}}{r} \sum_{\mu = 0}^\infty  \left( \frac{\lambda t}{\hbar}\right)^{2\mu}
  \frac{(-1)^\mu \Gamma\left(\mu + 1 \right)}{(2\mu)! \, A_+^{\mu + \frac{1}{2}}\Gamma(1)}   M_{\mu + \frac{1}{2},0}\left(\frac{r^2}{4A_+} \right),
\end{split}
\end{equation}
where $M$ is the Whittaker $M$-function. Then
\begin{equation}
\begin{split}
\label{eqn:functionD2r2}
q_\r(t) 
 &=     \frac{\nu \nu_0  \exp{\left(\frac{\rmi}{\hbar}\bar{\mu} t\right)}  }{\pi\sqrt{2\pi}} 2\pi 
 \frac{\exp{\left(-\frac{r^2}{8 A_+} \right)}}{r} \sum_{\mu = 0}^\infty  \left( \frac{\lambda t}{\hbar}\right)^{2\mu}
  \frac{(-1)^\mu \Gamma\left(\mu + 1 \right)}{(2\mu)! \, A_+^{\mu + \frac{1}{2}}}   M_{\mu + \frac{1}{2},0}\left(\frac{r^2}{4A_+} \right).
\end{split}
\end{equation}

\end{document}